\documentclass[a4paper,12pt]{article}

\usepackage[latin1]{inputenc}
\usepackage[dvips]{graphicx}
\usepackage[dvips]{color}
\usepackage{epsfig}
\usepackage{amsmath}
\usepackage{amssymb}
\usepackage{amscd}
\usepackage{theorem}
\usepackage{ifthen}
\usepackage[dvips]{graphicx}
\usepackage{dsfont}
\usepackage[all]{xy}
\usepackage{enumerate}

\numberwithin{equation}{section}

\newcommand{\ha}{{\textstyle \frac{1}{2}}}
\newcommand {\Zi}{Z_{\infty}}
\newcommand {\ri}{\right}

\newcommand {\Rmin}{R_{\rm min}}
\newcommand{\pr}{{\bf{Proof: }}}

\newcommand{\eop}{\hfill $\Box$ \\}
\newcommand{\idty}{{\rm 1\mskip-4mu l}}
\newcommand{\er}{\beta_E}
\newcommand{\aer}{\tilde{\beta}_E}
\theorembodyfont{\rm}
{\theorembodyfont{\normalfont}\newtheorem{theo}{Theorem}[section]}
{\theorembodyfont{\normalfont}\newtheorem{rem}[theo]{Remark}}
{\theorembodyfont{\normalfont}\newtheorem{lem}[theo]{Lemma}}
{\theorembodyfont{\normalfont}\newtheorem{prop}[theo]{Proposition}}
{\theorembodyfont{\normalfont}\newtheorem{defi}[theo]{Definition}}
\newcommand{\qmbox}[1]{\quad\mbox{#1}\quad}
\renewcommand {\l}{\left}
\newcommand {\LA}{\left\langle}
\newcommand {\RA}{\right\rangle}
\newcommand{\beq}{\begin{equation}}
\newcommand{\eeq}{\end{equation}}
\newcommand{\Leq}[1]{\label{#1}\eeq}
\newcommand{\beqn}{\begin{eqnarray}}
\newcommand{\eeqn}{\end{eqnarray}}
\newcommand{\beqno}{\begin{eqnarray*}}
\newcommand{\eeqno}{\end{eqnarray*}}

\newcommand {\bN}{{\mathbb N}}
\newcommand {\bR}{{\mathbb R}}
\newcommand {\bZ}{{\mathbb Z}}
\newcommand {\eh}{{\textstyle \frac{1}{2}}}
\newcommand {\q}{{\vec{q}}}
\newcommand {\p}{{\vec{p}}}

\newcommand {\pqs}{(\p_{0},\q_{0})}

\newcommand {\y}{{\vec{y}\,}}
\newcommand {\z}{{\vec{z}\,}}

\newcommand {\Rvir}{R_{\rm vir}}       

\newcommand {\htop}{{h_{\rm top}}}

\newcommand {\Opm}{\Omega^{\pm}}

\newcommand {\vep}{\varepsilon}

\newcommand {\Eth}{{E_{\rm th}}}

\newcommand{\rstr}{{\upharpoonright}}
\newcommand{\cC}{{\cal C}} 
\newcommand{\cM}{{\cal M}}
\newcommand{\cO}{{\cal O}} 

\newcommand {\Po}{{\cal P}_E} 

\newcommand{\NN}{\nonumber}

\newcommand{\bem}{\l(\! \begin{array}}
\newcommand{\eem}{\end{array}\!\ri)}
\newcommand{\bsm}{\left(\begin{smallmatrix}} 
\newcommand{\esm}{\end{smallmatrix}\right)}  

\begin{document}
\title {The Escape Rate of a Molecule}
\author{Andreas Knauf\thanks{Department Mathematik,
Universit\"{a}t Erlangen-N\"{u}rnberg,
Bismarckstr.\ $1 \eh$, D--91054 Erlangen, Germany.
e-mail: knauf@mi.uni-erlangen.de}\hspace{2cm} Markus Krapf}
\date{May 02, 2008}
\maketitle
\begin{abstract}\noindent
We show existence and give an implicit formula for the escape rate of 
the $n$-centre problem of
celestial mechanics for high energies. 
Furthermore we give precise computable estimates of this rate.
This  exponential decay rate plays an important role
especially in semiclassical scattering theory of $n$-atomic molecules.
Our result shows that the diameter of a molecule is measurable
in a (classical) high-energy scattering experiment.
\end{abstract}
%
\section{Introduction and Statement of Results}\label{sec:def_n-centre}
%
The {\em $n$-centre problem} in
three dimensions is given by $n$ {\em nuclei} with {\em charges}  $Z_1,\ldots,
Z_n\in\mathbb{R}\setminus\{0\}$ fixed at positions
$\vec{q}_1,\ldots,\vec{q}_n\in \mathbb{R}^3$.
We assume that the nuclei are
in \emph{general position}, which  means that no three $\vec{q}_k$ 
lie on one line.
The $n$-atomic molecule generates a
Coulombic potential  on the {\em configuration space} 
$\hat{M}:=\mathbb{R}^3\setminus\{\vec{q}_1,\ldots, \vec{q}_n\}$ :
\begin{defi}\label{def:Coulomb}
A smooth {\em potential} $V:\hat{M}\to\bR$ is called {\em Coulombic} if
\begin{enumerate}
\item $V$ has the form

$$V(\vec{q\,}) =
\sum_{k=1}^{n}\frac{-Z_{k}}{\|\vec{q}-\vec{q}_{k}\|}+ W(\vec{q\,}) \qquad (\vec{q\,}\in\hat{M})$$
with $W:\mathbb{R}^3 \rightarrow \mathbb{R}$ smooth.

\item 
The potential vanishes at infinity, {\em i.e.}
$\lim_{\|\vec{q\,}\|\rightarrow\infty}V(\vec{q\,})=0$, and its difference to a
Coulomb potential is of \emph{short range}. I.e.\ there exists $\Zi\in\mathbb{R}$,
called the \emph{asymptotic charge},
$\varepsilon\in(0,1]$ and
\[\Rmin>2\max(\|\vec{q}_1\|,\ldots,\|\vec{q}_n\|)\]
such that for some $C_1>0$
$$\left\|\nabla V(\q) - \Zi\frac{\q}{\|\vec{q\,}\|^{3}}\right\| <
\frac{C_1\;\Rmin}{\|\q\|^{2+\varepsilon}}\qquad(\|\vec{q\,}\|\geq \Rmin)$$and
$\|\nabla V(\q_{1}) - \nabla V(\q_{2})\|<
C_1\frac{\|\q_{1} - \q_{2}\|}{\min(\|\q_{1}\|,\|\q_{2}\|)^{2+\varepsilon}}
\qquad(\|\q_1\|,\|\q_2\|\geq \Rmin)$.
\end{enumerate}
\end{defi}
\begin{rem} \label{rem:TF}
\begin{enumerate}
\item 
In \cite{knauf} the high energy dynamics in Coulombic potentials was
analyzed, using symbolic dynamics. The results are used here to
calculate the escape rate, a quantity measurable in scattering
experiments. For semiclassical aspects of the model, see \cite{CJK},
for topological methods also applicable to non-singular potentials, 
see \cite{KK2}. 

Derezi\'nski and G\'erard \cite{derezinski} is 
used as a general reference for 
scattering theory. 

Narnhofer
analyzed time delay for short range potentials in \cite{narnhofer}.
\item 
In the context of celestial mechanics, $V$ is sum of the
singular Kepler potentials, with $Z_i>0$ interpreted as masses, and $W=0$.

For electrostatic potentials the charges may be positive and negative,
i.e.\ the force can be attractive as well as repulsive.
The scattering of a classical electron by a molecule can be well
modeled in this setting by positive charges of the nuclei, i.e.\
$Z_i>0$, and an additional smooth shielding (electronic) potential
$W$, say of Thomas-Fermi type, such that, up to a Coulombic term 
$Z_\infty/\|\q\|$ given by the net charge of the
molecule, the resulting potential is of short range, see
\cite{CJK}. 

Note that for $W=0$ we have $Z_\infty=\sum_{i=1}^nZ_i$.

Both for $W=0$ and in the Thomas-Fermi case one may take $\vep=1$ 
in Def.\ \ref{def:Coulomb}.
\end{enumerate}
\end{rem}
The Hamilton function $\hat{H}:T^*\hat{M}\rightarrow \mathbb{R}$
on the phase space $T^*\hat{M}\simeq\mathbb{R}^3\times \hat{M}$
is given by
\beq
\hat{H}(\vec{p\,},\vec{q\,}\,):=\ha\|\vec{p\,}\|^{\;2}+V(\vec{q\,}).
\Leq{s}

For $n=1$ centres this generalizes the Kepler problem. For
large energies no bounded orbits exist, for which reason the time delay is
bounded. So we assume $n\geq 2$ in the following.

Due to collision orbits for nuclei with positive charge 
(respectively mass, depending on the interpretation)
the Hamiltonian flow generated by (\ref{s}) is incomplete.
As we are interested in time-related quantities like time delay and escape
rate  we use a regularization method which -- 
unlike the so-called Kustaanheimo-Stiefel transform -- 
does not involve a time reparametrization. 
In Section 5 of \cite{knauf} such a regularization
is done by phase space extension of $T^*\hat{M}$.

After regularization we get a smooth Hamiltonian system
$(P,\omega, H)$ with a six-dimensional smooth manifold $P\supset
T^*\hat{M}$, $H\in\mathcal{C}^\infty(P,\mathbb{R})$ with
$H\rstr_{T^*\hat{M}}=\hat{H}$ and a symplectic two-form
$\omega\in\Omega_2(P)$ such that $\omega\rstr_{T^*\hat{M}}=\omega_0$,
with $\omega_0=\sum_{i=1}^{3}dq_i\wedge dp_i$ the canonical
symplectic form on $T^*\hat{M}$. 

This Hamiltonian system generates a smooth complete Hamilton flow
$\Phi:\mathbb{R}\times P\rightarrow P$.
Although for collisions the momentum $\p$ diverges, for
simplicity we write the flow in the forms
\[\big(\vec{p\,}(t,x),\vec{q\,}(t,x)\big):=\Phi_t(x):=\Phi(t,x)\qquad
(x\in P,\ t\in\mathbb{R}).\]
%
\subsection{Classification of States in Phase Space}
%
We are dealing with Hamiltonian dynamics for which the \emph{energy} (i.e.\ the value of the Hamilton function) is
conserved. So for a given energy $E\in H(P)$ the dynamics is
confined on the \emph{energy surface} $\Sigma_E:=H^{-1}(E)\subset
P$. For large $E$ this is a smooth manifold of dimension five.

We now classify the points in the (extended) phase
space $P$ by their asymptotic behaviour:
\begin{itemize}
\item states   $b^\pm:=\{x\in P: \limsup\limits_{t\rightarrow\pm\infty}\|\vec{q\,}(t,x)\|<\infty\}$,
  \emph{bounded in the future} respectively {\em past}, 
  and the \emph{bounded states}  $b:=b^+\cap b^-$
\item  states  $s^{\pm}:=P\setminus b^{\pm}$, \emph{scattered in the future/past} and the
  \emph{scattered states} $s:=s^+\cap s^-$ 
\item  states $t^{\pm}:= s^{\mp}\setminus s^{\pm}=s^{\mp}\cap b^{\pm}$ 
 \emph{trapped in the   future/past} and  \\
  the \emph{trapped states}
  $t:=s^+\Delta s^-=(b^+\cap s^-)\cup (b^-\cap s^+)=t^+\cup t^-$.
\item with a subscript $E$ we denote the corresponding  sets restricted to the
  energy surface $\Sigma_E$, i.e.\ $b_E:=b\cap \Sigma_E$.
\end{itemize}
The assumption that $V$ is Coulombic gives rise to the {\em virial
inequality}
\beq
\frac{d}{dt} \LA\q(t),\p(t)\RA =
2\Big(E-V\big(\q(t)\big)\Big)+\LA\q(t),\nabla V\big(\q(t)\big)\RA\ge E > \Eth \ge 0. 
\Leq{VI}
This is valid for $H(x)=E$  and the trajectory
$t\mapsto \vec{q\,}(t):=\vec{q\,}(t,x)$ 
outside an \emph{interaction zone} defined by a
\emph{virial radius} $\Rvir$
\beq
\mathcal{IZ}(E):=\left\{\vec{q\,}\in \mathbb{R}^3:\|\vec{q\,}\,\|\leq 
\Rvir(E)\right\} \qquad (E>0).
\Leq{ss}
So a trajectory leaving the
interaction zone at some time $t_0$ will move away from the
origin for all future times $t>t_0$, and for any scattered state 
$x\in s^\pm_E$ also
$\lim_{t\rightarrow\pm\infty} \|\vec{q\,}(t,x)\|=\infty$.

The function $E\mapsto \Rvir(E)$ can be chosen to be continuous and
non-increasing in $E$, and as we consider high energies, i.e.\
energies above an \emph{energy threshold} $\Eth>0$,
we can assume
a energy-independent virial radius $\Rvir:=\Rvir(\Eth)$ and
energy-independent interaction zone
$\mathcal{IZ}:=\mathcal{IZ}(\Eth)$,
see \cite{knauf}, p.\ 11 for details. In the course of the article
several other lower bounds on the constant $\Eth$ will arise.

We consider the asymptotic behaviour of the flow
$\Phi_t:P\rightarrow P, x\mapsto (\vec{p\,}(t,x),\vec{q\,}(t,x))\in P$ by
defining for $E>0$ and a point $x\in s_E^{\pm}$ the
\emph{asymptotic velocities, directions} and
\emph{impact parameters}
\[\vec{p}_\pm(x):=\lim_{t\to\pm\infty}\p(t,x)\qmbox{,}
\hat{p}^{\pm}(x):=\frac{\vec{p\,}^{\,\pm}(x)}{\sqrt{2E}}\in S^2\]
\[\vec{q\,}_{\perp}^{\;\pm}(x):=
\lim\limits_{t\rightarrow \pm\infty}\big(\vec{q\,}(t,x)-\langle
\vec{q\,}(t,x),\hat{p}^{\pm}(x)\rangle
\cdot\hat{p}^{\pm}(x)\big).\] 
These are $\Phi_t$-invariant and depend continuously on the point $x$,
see Theorem 6.5 of \cite{knauf}.

Noting that $\hat{p}^{\pm}(x)\perp \vec{q\,}_{\perp}^{\;\pm}(x)$, 
we define the continuous \emph{asymptotic maps}
\beq\label{eq:asymptotic_map}
A^{\pm}_E:s_E^{\pm}\rightarrow T^*S^2\qmbox{,}x\mapsto
(\hat{p}^{\pm}(x),\vec{q\,}_\perp^{\pm}(x)).
\eeq
We denote the canonical symplectic two-form on the cotangent bundle
$T^*S^2$ of the sphere by $\omega_0\in \Omega_2(T^*S^2)$, and use
the volume four-forms  
$$\Omega_E:= E\, \omega_0\wedge\omega_0\in \Omega_4(T^*S^2)\qquad (E>\Eth).$$

The sets $D_E^{\pm}:=A_E^{\pm}(s_E)\subset T^*S^2$ represent the possible
asymptotic data in the corresponding time direction, for a given energy
$E$.

The \emph{asymptotic scattering map of energy $E$}
\beq
AS_E:D_E^-\rightarrow
D_E^+ \qmbox{,} AS_E:=A_E^+\circ
(A_E^{-})^{-1}
\label{pg:asymptotic_scattering_map}
\eeq
maps the initial asymptotic data $(\hat{p}^-,\vec{q\,}_{\perp}^{\,-})\in D_E^-$
to the final asymptotic data $(\hat{p}^+,\vec{q\,}_{\perp}^{\,+})\in D_E^+$.

\begin{rem}\label{rem:asymptotic_completeness}
Note that $A^\pm_E(s_E^\pm)=T^*S^2$ but
$T^*S^2\setminus D_E^{\pm}\not=\emptyset$ if the sets
$s_E^{\pm}\setminus s_E = t_E^{\mp}$
of past/future trapped orbits are not empty. The \emph{asymptotic
completeness} of the $n$-centre problem, implies that $D_E^\pm$ is of full
measure with respect to the canonical volume form $\Omega_E$ on $T^*S^2$,
see Corollary 6.4 in \cite{knauf}.
\end{rem}

Since the asymptotic maps $A_E^{\pm}$ on $s_E^{\pm}$ are
$\Phi_t$-invariant, the asymptotic scattering map $AS_E$
carries no information about time-related quantities.
%
\subsection{M{\o}ller Transformation}
%
Far from the origin in configuration space, the $n$-centre problem is well 
approximated by the Kepler problem, given by the phase space
$\hat{P}_{\infty}:=T^*(\mathbb{R}^3\setminus\{0\})$ and Hamilton function
$$\hat{H}_{\infty}:\hat{P}_\infty\rightarrow \mathbb{R}\qmbox{,}
\hat{H}_{\infty}(\vec{p}_\infty,\vec{q}_\infty):=
\ha \|\vec{p}_\infty\|^{\,2}-\frac{Z_\infty}{\|\vec{q}_\infty\|}.$$
Being for $Z_\infty>0$ a special case of (\ref{s}), it can 
always be regularized to yield a smooth complete flow
\beq
\Phi_t^\infty:P_\infty\rightarrow P_\infty\qmbox{,}
x_\infty\mapsto \big(\vec{p}_\infty(t,x_\infty),\vec{q}_\infty(t,x_\infty)\big)
\Leq{flow:infty}
of the Kepler problem, with the extended Hamilton function
$H_\infty\in\mathcal{C}^\infty(P_\infty,\mathbb{R})$.

The scattering states  of the flow $\Phi^\infty$ with a
non-vanishing asymptotic momentum form the set
$$P_{\infty,+}:=\{x\in P_\infty: H_\infty(x)>0\},$$ consisting of
$\Phi^\infty$-orbits projecting  to Kepler hyperbolae (resp.\ straight lines
for $Z_\infty=0$) in configuration space.

Scattering theory in general deals with the comparison of two
dynamics, in this case the dynamics of $(P,\Phi_t)$ and
$(P_\infty,\Phi_t^\infty)$. This is done by the \emph{M{\o}ller
transformations} 
\beq\label{eq:moeller_transform}
\Omega^{\pm}:P_{\infty,+}\rightarrow s^{\pm} \qmbox{,}
\Omega^\pm:=\lim_{t\rightarrow\pm\infty}\Phi_{-t}\circ\Phi_t^\infty.
\eeq
Here we have omitted the identification of $P$ with $P_{\infty}$
outside a region near the singularities.

In  \cite{knauf}, Sect.\ 6 it was shown that the M{\o}ller transformations $\Omega^{\pm}$
exist point-wisely and are measure-preserving homeomorphisms, and
if the partial derivatives of $V$ decay at infinity like
$$
\partial^{\beta}_q  \left( V(\q) + \frac{Z_\infty}{\|\vec{q\,}\|} \right)
\ \stackrel{ \vec{q} \rightarrow \infty }{=}\
{\cal O}\left(\|\vec{q\,}\|^{-|\beta|-1-\varepsilon}\right)
\qquad  (\beta\in \mathbb{N}_{0}^3)
$$
for some $0<\varepsilon\leq 1$, then the M{\o}ller transformations
are $\mathcal{C}^\infty$- symplectomorphisms.\\
Similar statements hold for the asymptotic scattering map, defined
in (\ref{pg:asymptotic_scattering_map}).
Like in Remark \ref{rem:TF}, 
both for $W=0$ and in the Thomas-Fermi case one may take $\vep=1$.

We denote by $\Omega_*^\pm:s^{\pm}\rightarrow P_{\infty,+}$ the inverse of $\Omega^\pm$.

 From this  follows in particular that
$\Omega^{\pm}(P_{\infty,+})=s^{\pm}$, i.e.\ given a Kepler hyperbola
there exists a unique scattered orbit of the $n$-centre problem which is
asymptotic to this hyperbola in the time direction described by the
sign. Conversely any one-sided scattered orbit is one-sided
asymptotic to a unique Kepler hyperbola.

The choice of the Kepler problem as the ``comparison  dynamics'' for
defining the M{\o}ller transformation is
justified by the existence of the limit in (\ref{eq:moeller_transform}).
%
\subsection{Time Delay and Escape Rate}
%
Next we define the \emph{time delay} $\tau(x)$
for a point $x$ belonging to a scattered orbit by comparison with
the Kepler dynamics ($\Theta$ denoting the Heaviside step function):
\beq
\tau(x) := \lim_{R\rightarrow\infty}\int_\mathbb{R}
\Theta\big(R-||\vec{q\,}\,(t,x)||\big)-\ha\Big[\Theta\big(R-||\vec{q}_{\infty}(t,\Omega^+_*(x))||\big)
+\Theta(R-||\vec{q}_{\infty}(t,\Omega_*^-(x))||\big)\Big]dt.
\Leq{eq:time_delay}
As $\tau$ is $\Phi$-invariant, the time delay $\tau(x)$ only
depends  on the asymptotic data $(\hat{p}^+,\vec{q\,}^+_{\perp})\in
T^*S^2$ (respectively $(\hat{p}^-,\vec{q\,}^-_{\perp})\in T^*S^2$) of
$x$. So for $E>0$ we define $\tau_E:=\tau\rstr_{\Sigma_E}$ and the
\emph{asymptotic time delay}

\beq
\tau_E^{+}:T^*S^2\rightarrow \mathbb{R}\cup\{\infty\}\qmbox{,}
\tau_E^{+}(\hat{p},\vec{q\,}_\perp):=\left\{\begin{array}{ccc}
\tau_E\circ \big(A_E^{+}\big)^{-1}(\hat{p},\vec{q\,}_{\perp})&
\mbox{if } (\hat{p},\vec{q\,}_{\perp})\in D_E^{+} \\
\infty & \mbox{else} \end{array}\right.
\Leq{eq:time_delay_as}
(remember that $D_E^+ = A_E^+ (s_E)$).
\beq
\kappa^{\infty}_E(t):=\int_{T^*S^2}
\idty_{\{\tau_E^+(x)\geq t\}}\Omega_E(x)\qquad (t>0)
\Leq{eq:int_time_delay}
denotes the $\Omega_E$-volume of the set of
asymptotic data with a time delay greater or equal than
a given time delay $t>0$.
Clearly  $\kappa_E^\infty(t)$ is monotone decreasing in $t$.

As the asymptotic scattering map
(\ref{pg:asymptotic_scattering_map}) is measure-preserving,
this quantity would not change if one
would use $A_E^{-}$ instead of $A_E^{+}$ in
Def.\ (\ref{eq:time_delay_as}).
\begin{rem}\label{rem:bound}
The passage from the time delay to the asymptotic time delay is
motivated by measure theoretical properties of these maps:
because of $\Phi$-invariance of $\tau$, for any time interval 
$I\subset \mathbb{R}$ the 
Liouville measure of the set $\{x\in s_E : \tau_E(x)\in I\}$
of scattered states is either zero or infinite, whereas 
(\ref{eq:int_time_delay}) turns out to be finite for $t>0$.
\end{rem}
Although our main interest lies in the analysis of orbits with large
time delay, we also show
\begin{enumerate}
\item
that the time delay is bounded below
on the energy surface, 
\item
that the volume of orbits with a given  
positive time delay is bounded, 
\item
and that complicated dynamics of
scattering 
orbits only occur for large time delay.
\end{enumerate}
More precisely we have the following results:
\begin{prop}\label{prop:small}
There exist constants $C_1, C_2, C_3>0$ such that for all energies $E>\Eth$
\begin{enumerate}
\item
$\tau_E>- C_2/E^{1/2}$
\item
with $\vep\in(0,1]$ from Def. \ref{def:Coulomb} we have 
$\kappa_E^\infty(t)< C_1 t^{-2/\vep}E^{1-3/\vep}\qquad
\big(t\in(0,C_3/E^{3/2})\big)$
\item
the orbit through $x$ intersects the interaction zone if $
\tau_E(x)\in\big[C_3/E^{3/2},\infty\big)$.
\end{enumerate}
\end{prop}
The proof of Proposition \ref{prop:small} is in the Appendix.\\[2mm]
For {\em large time delay} the \emph{escape rate} $\er$ is defined as the
exponential decay rate of $\kappa^{\infty}_E(t)$, i.e.\
\beq
\er:=-\lim\limits_{t\rightarrow\infty}\frac{1}{t}\ln\big(\kappa_E^\infty(t)\big)
\qquad (E>\Eth)
\Leq{eq:de:escape_rate}
in the case of existence.

Now we are ready to state our main result. To ease the notation,
for real valued functions $f,g$ we write $f\asymp g$
(or sloppily $f(t)\asymp g(t)$)
if there exist constants $C_1\geq 1$, $C_2>0$ such that
$C_1^{-1} g(t)\leq f(t)\leq C_1 g(t)$ for all $t\geq C_2$. 
%
\subsection{A Matrix Perron-Frobenius Problem}
%
We now set up a finite matrix problem which we show to approximate 
the escape rate (\ref{eq:de:escape_rate}) with optimal precision (see Remark
\ref{rem:precise}).
This then allows to compute the escape rate very precisely, using only
a few parameters of the model. 
Symbolic dynamics will be based on the {\em alphabet}
\beq
\mathcal{A}:=\{(i,j): i,j=1,\ldots, n\mbox{ with } i\not=j\}.
\Leq{def:A}
For 
$(k_0,k_1)=\big((i,j),(k,l)\big)\in \mathcal{A}\times\mathcal{A}$ with $j=k$
we define the charge $Z_{k_0,k_1} := Z_j$  and, for 
distance $d_{i,j}:=\|\vec{q}_i-\vec{q}_j\|$, 
the mean distance $\overline{d}_{k_0,k_1}:=\eh(d_{i,j}+d_{j,l})$.
We set
\beq
f\big(k_0,k_1\big):=
\frac{2d_{i,j}\cos^2(\eh\alpha(i,j,l))}{-Z_{j}},
\Leq{def:f}
$\alpha(i,j,l)$ denoting the angle between the vectors 
$\vec{q}_i-\vec{q}_j$ and $\vec{q}_l-\vec{q}_j$. Then, with
\beq
\widetilde{T}_E(k_0,k_1):=\frac{\overline{d}_{k_{0},k_{1}}}{\sqrt{2E}}-\frac{Z_{k_0,k_1}\ln(E)}{(2E)^{3/2}}
\qmbox{and} 
\widetilde{F}_E(k_0,k_1):=2\ln\left(\frac{2E|f(k_0,k_1)|}{d_{k_0}}\right)
\Leq{eq:approximations}
playing the role of approximate Poincar\'{e} time respectively unstable Jacobian, 
we define the weighted transfer matrix 
(consult Baladi \cite{baladi} for the subject of  transfer operators)
\beq
\left(\mathcal{M}_E\big(\beta\big)\right)_{k_0,k_1}:=\left\{
   \begin{array}{cc}
  \exp\left(-\widetilde{F}_E(k_0,k_1)+\beta\;\widetilde{T}_E(k_0,k_1)\right) & \mbox{if } j=k \\
 0 & \mbox{else}
   \end{array}
\right..
\Leq{M:E}
Note that the definition of $\mathcal{M}_E$ 
only involves the energy $E$, the positions $\vec{q}_i$
and the charges $Z_i$ of the $n$ nuclei
and is independent of the potential $W$.

The Perron-Frobenius eigenvalue $\lambda_{\rm PF}>0$ of $\mathcal{M}_E$
now depends on $E$ and $\beta$, and will be shown to have 
a unique solution $\tilde{\beta}_E$ of
\[\lambda_{\rm PF}(\beta,E)=1\qquad (E>\Eth).\]
By $d_{\max}$ we denote the maximal mutual 
distance of the nuclei, that is the diameter of the molecule.\\
 
\begin{figure}
\centerline{
\epsfig{file=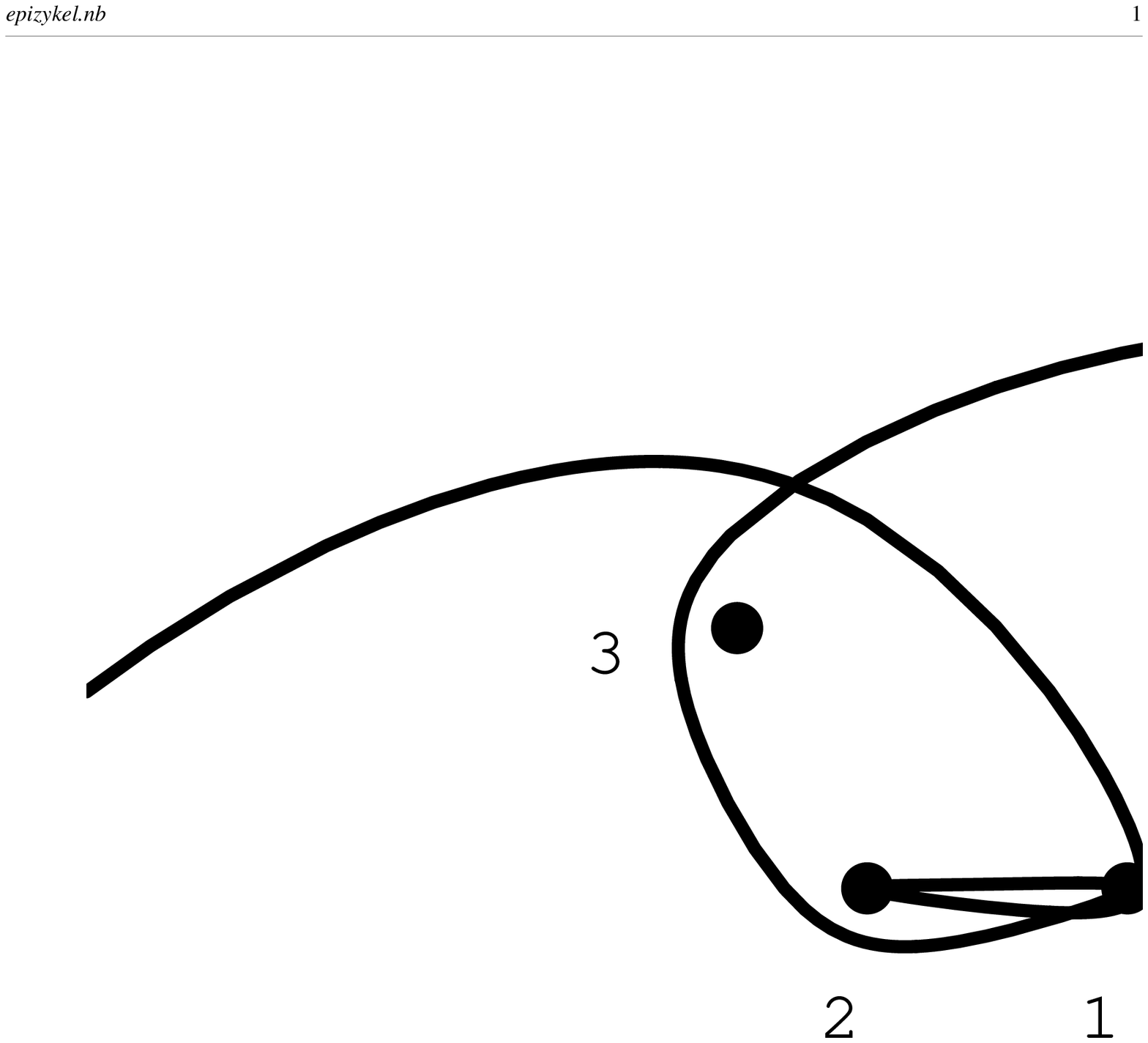,height=5cm,bbllx=80,bblly=300,bburx=590,bbury=650, 
clip=}%
\hspace*{20mm}
\epsfig{file=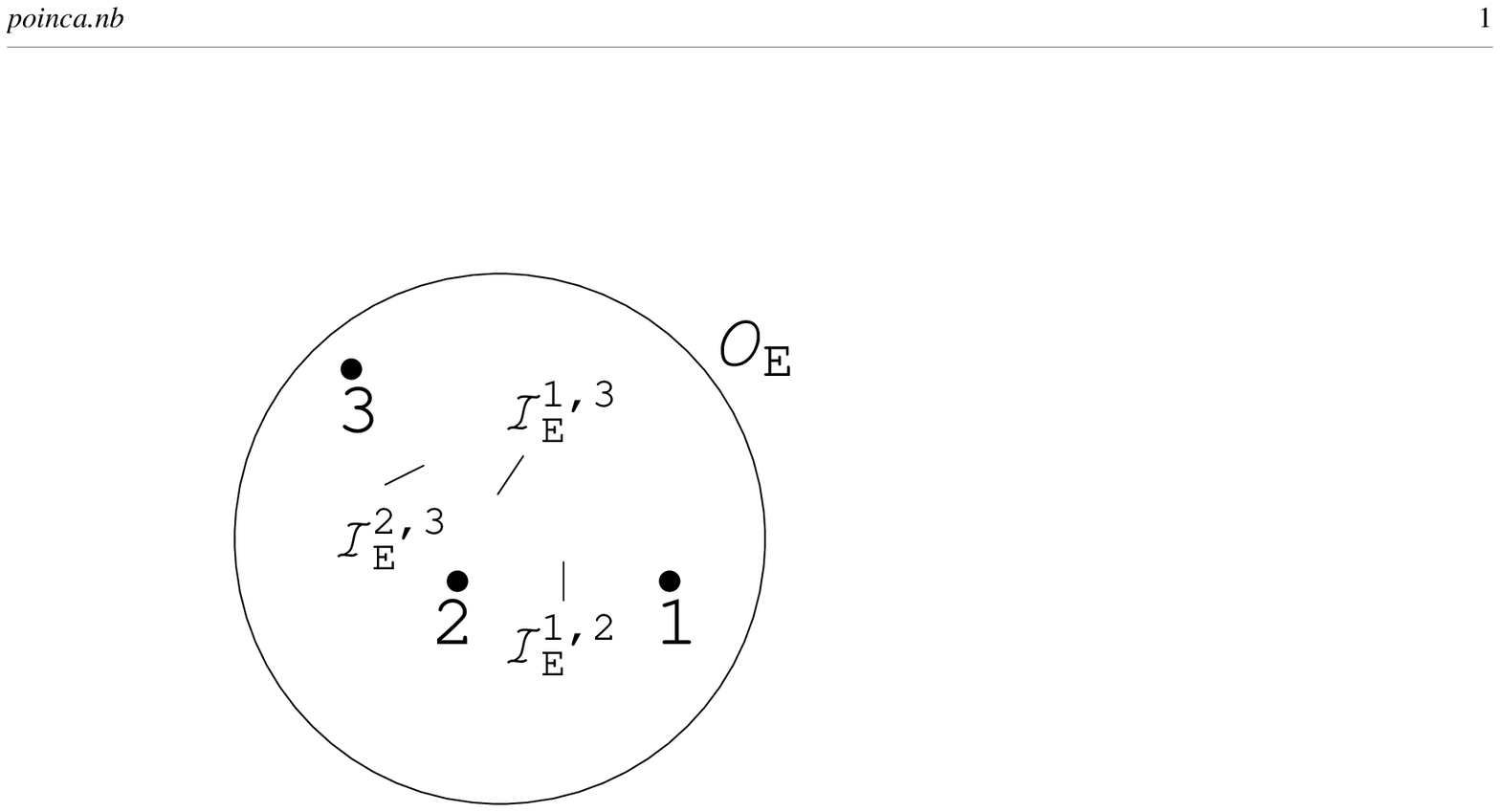,height=5cm,bbllx=110,bblly=480,bburx=330,bbury=680, 
clip=}%
}
\caption{Left: Scattering orbit for $n=3$ centres, 
with symbol sequence $1,2,1,2,3$. 
Right: Poincar\'{e} surfaces projected to the plane in
configuration space, containing the three centres}
\label{fig:only}
\end{figure}

\noindent
{\bf Main Theorem}:
Let $V$ be a Coulombic $n$-centre potential, $n\geq 2$, and the energy
$E>\Eth$.
Then for the escape rate $\beta_E$
(defined in Eq.\ (\ref{eq:de:escape_rate})) it holds:
\begin{enumerate}
\item[(i)] $\beta_E$ exists, even more 
$\kappa_E^\infty(t)\asymp e^{-\er t}$.\\
$\er$ is given implicitly by a Perron-Frobenius problem.
\item[(ii)] The escape rate $\er$ is asymptotic to 
$\frac{2\sqrt{2E}\ln E}{d_{\max}}$. It is
approximated by $\tilde{\beta}_E$,
with relative error of order $\mathcal{O}(1/E)$.
\end{enumerate}
\begin{rem}\label{rem:precise}
As a $W$-independent estimate,
the $\cO(1/E)$ estimate of (ii) is optimal. This can be seen by adding to $V$ a
cut off function $W\in C^\infty_c (\mathbb{R}^3)$
which equals a constant $C$
in the interaction zone ${\mathcal IZ}$, see (\ref{ss}). Thus the  dynamics in
$\Sigma_E$ over ${\mathcal IZ}$ equals the one without $W$ for
energy $E-C$.

We base our proof on the application of renewal theory, worked out by 
Lalley in \cite{lalley}, precisely determining the $E$-dependence of
all quantities. This is possible by applying a simple symbolic
dynamics from \cite{knauf}, see Figure \ref{fig:only}.
\end{rem}
%
\section{Proof of the Main Theorem}
%
While our estimates are optimal in their dependence on the energy $E$,
we will be somewhat vague in denoting most energy-independent constants by
$C$, without tracing back their mutual dependence.
We hope that this makes the following part more readable.
The interested reader, however, may consult \cite{knauf} to find
in many cases more explicit estimates.
%
\subsection{Proof of Part (i)}
%
Although the escape rate is defined in the realm of scattering theory,
the key of showing its existence and determining its value is to study the bounded states.
As the escape rate is a limit of large time delay, it is natural
that the trapped states play an important role. But
in our case for $E>\Eth$ the $\omega$-limit set of
the trapped states
equals the bounded states (i.e.\ $\omega(t_E)=b_E$).

\subsubsection{Symbolic Dynamics}
%
The set of non-wandering points of
the flow $\Phi$ on the energy surface $\Sigma_E$ equals $b_E$,
the subset of bounded states.
Moreover, for high enough energies $E>\Eth$ \ $b_E$  is a hyperbolic set 
so that the flow
${\Phi_t}\rstr_{ \Sigma _E}$ satisfies Axiom A (see \cite{knauf}, Thm. 12.8).
This allows, by using Poincar\'e sections, to model the
time-discretized dynamics ${\Phi_t}\rstr_{b_E}$ with
symbolic dynamics given by a two-sided shift space $(X,\sigma)$ of finite type.
The left shift $\sigma$ on $X$ in conjugated to the \emph{Poincar\'e map}
$\mathcal{P}_E$, restricted to the bounded states on the Poincar\'e surfaces.
\begin{lem}[Theorem 12.8 of \cite{knauf}]\label{theo:symb_dyn_bounded}
For energies $E>\Eth$ the flow ${\Phi_t}\rstr_{b_E}$ of the $n$-centre
problem is conjugated to a suspended flow
$(X^E,\sigma_t^E)$
by a H\"older continuous homeomorphism $\mathcal{X}^E:X^E\rightarrow b_E$.
\hfill $\Box$
\end{lem}
\begin{rem}
H\"{o}lder continuity is defined by a choice of Riemannian
metric on the Poincar\'e sections, denoted by $d_{\mathcal{I}_E}$
and a metric on the shift space, see (\ref{me:me}) below.

The roof function is the pull-back $T_E:X\rightarrow\mathbb{R}^+$ of the 
Poincar\'e time, see (\ref{P:T}) below.
In particular it is H\"{o}lder continuous on the shift space $X$.

 From Theorem 19.1.6 and its Corollary 19.1.13 of  Katok and Hasselblatt \cite{katok}
it follows that the logarithm of the unstable Jacobian, denoted by $F_E$ (see (\ref{FE}) below)
is also a H\"{o}lder continuous function on the intersection
of the Poincar\'e surfaces with the bounded states.
\end{rem}
This conjugacy is established by introducing Poincar\'e surfaces
labeled by the {alphabet} $\mathcal{A}$ from (\ref{def:A}).
For $C>0$ the hypersurfaces in the energy
shell $\Sigma_E$, labelled by $(i,j)\in \mathcal{A}$
\beqn\label{eq:surfaces}
\mathcal{I}^{i,j}_E&:=&\Big\{(\vec{p\,},\vec{q\,})\in \Sigma_E:
\langle \vec{q\,}-\vec{m}_{i,j},\hat{q}_{i,j}\rangle=0,\,
|\vec{q\,}-\vec{m}_{i,j}|< \frac{C\, d_{i,j}}{2\,E},
\nonumber \\
& & \hspace*{28mm} \langle \vec{p\,},\hat{q}_{i,j}\rangle>0,\; \Big|\frac{\vec{p\,}}{\sqrt{2E}}\times \hat{q}_{i,j}\Big|
<2C\, E^{-1}\Big\} ,\qquad 
\eeqn
are located in configuration space near the midpoint
$\vec{m}_{i,j}:=\ha(\vec{q}_i+\vec{q}_j)$ between the centre $i$ and
$j$, and are perpendicular to the direction
$\hat{q}_{i,j}:=(\vec{q}_j-\vec{q_i})/d_{i,j}$ (with
$d_{i,j}=\|\vec{q}_i-\vec{q_j}\|$), see Figure \ref{fig:only}.

For an appropriate constant $C>0$ these
are Poincar\'e surfaces for all $E>\Eth$, i.e.\ they are transversal to the flow 
$\Phi_t\rstr_{\Sigma_E}$.

We denote the disjoint union of these \emph{inner Poincar\'e surfaces}
by 
$\mathcal{I}_E:=\bigcup_{({i,j})\in \mathcal{A}}\mathcal{I}_E^{i,j}$.

$\mathcal{D}_E:= \big\{x=(\vec{p\,},\vec{q\,})\in \Sigma_E:\
\|\vec{q\,}\|\leq \Rvir\big\}$
denotes the part of the energy surface lying over the
interaction zone (\ref{ss}).
The two submanifolds of the boundary
\beq\label{eq:boundary_pm}
\mathcal{O}_E^{\pm}:=\{(\vec{p\,},\vec{q\,})\in \Sigma_E: \| \vec{q\,}\,\|
= \Rvir ,\; \pm \langle \vec{q\,},\vec{p\,}\rangle >0 \}
\subset \partial \mathcal{D}_E,
\eeq
consisting of states
leaving resp.\ entering the interaction zone, are 
transversal to the flow,
too and called {\em outer Poincar\'e surfaces}. We 
use their disjoint unions
$$\mathcal{O}_E:=\mathcal{O}_E^{+}\cup\mathcal{O}_E^{-}\qmbox{,}\mathcal{H}_E^\pm
:=\mathcal{I}_E\cup\mathcal{O}_E^{\pm}\qquad\mbox{and}\qquad \mathcal{H}_E:=\mathcal{I}_E\cup\mathcal{O}_E.$$
The \emph{Poincar\'e map}
$\mathcal{P}_E:\mathcal{H}^-_E\rightarrow \mathcal{H}^+_E,\,
\mathcal{P}_E(x):=\Phi(T_E(x),x)$ with \emph{Poincar\'e time} is given by
\beq
T_E:\mathcal{H}_E\rightarrow [0,\infty)
\qmbox{,} T_E(x):=\left\{
\begin{array}{ccc}
0& \mbox{for}& x\in \mathcal{O}_E^+\\
\inf\left\{t>0:\Phi_t(x)\in\mathcal{H}_E^+\right\} & \mbox{for} & x\in\mathcal{H}_E^-
\end{array}\right..
\Leq{P:T}
The Poincar\'e map $\mathcal{P}_E$, restricted to
$\mathcal{I}_E\cap b_E$, gives rise to the \emph{two-sided shift space} $(X,\sigma)$
\beq\label{eq:shift_space_n-centre-problem}
X:=\l\{(\ldots,k_{-1},k_0,k_1,\ldots)\in \mathcal{A}^\mathbb{Z}:\; 
(k_{i+1})_1=(k_{i})_2\;\forall i\in\mathbb{Z}\ri\}.
 \eeq
Using similar definitions, with $X^+\subset \mathcal{A}^{\mathbb{N}_0}$ we denote 
the \emph{one-sided shift space} and with $X^*$ the set of (unindexed)
{\em words} in $X^+$ resp.\ in $X$.

With the metric
\beq
d(\underline{k},\underline{l}) := 
2^{-\sup\{i\in\mathbb{N}_0:\; k_j=l_j\;\forall |j|\leq i\}}
\label{me:me}
\eeq
the space $X$ resp.\ $X^+$ becomes a metric space.
According to Lemma 12.2 of \cite{knauf} we have symbolic dynamics
in the following sense:
there exists a H\"{o}lder
continuous homeomorphism
\beq
{\cal F}_E:X\rightarrow b_E\cap  \mathcal{I}_E,
\Leq{symb:homeo}
conjugating the shift on $X$ with 
the Poincar\'e map on the bounded orbits.
This allows us to consider functions on $b_E\cap  \mathcal{I}_E$
like Poincar\'{e} time $T_E$ as functions on $X$.

For a word $\underline{k}=(k_0,\ldots, k_m)\in X^*$ of length $m+1$ 
we denote the {\em cylinder over} $\underline{k}$ by
$$[\underline{k}]:=\{\underline{l}\in X: l_i=k_i\ \forall i=0,\ldots, m\}.$$
This corresponds to the open submanifold on the inner Poincar\'e surface 
\beq
\mathcal{I}_E(\underline{k}):=
\left\{ x\in \mathcal{I}_E: \mathcal{P}_E^ i(x)\in \mathcal{I}_E^{k_i},\ i=0,\ldots, m\right\}.
\Leq{I:E}
The canonical symplectic form $\omega$ on $P$, restricted to
$\mathcal{H}_E$ makes this four-dimensional Poincar\'e surface
a symplectic manifold.
We denote the restriction of the canonical volume four-form on
$\mathcal{H}_E$ by $\Omega_E$, like the volume on $T^*S^2$.


Next we define the  \emph{total inner Poincar\'e time}
$\tau_{\mathcal{I}_E}:\Sigma_E\rightarrow[0,\infty]
:=[0,\infty)\cup\{+\infty\}$, 
the time spent in the set $\mathcal{I}_E$ of inner Poincar\'e surfaces, by
$$\tau_{\mathcal{I}_E}(x):=\left\{
\begin{array}{cc}
\sup\{t\in\mathbb{R}:\Phi_t(x)\in\mathcal{I}_E\}-\inf\{t\in\mathbb{R}:\Phi_t(x)\in\mathcal{I}_E\}
& \mbox{ if }\mathcal{I}_E\cap \Phi({\mathbb{R}},x)\not=\emptyset\\
0 & \mbox {else}
\end{array}\right.$$
and consider the $\Omega_E$-volume
\beq
\kappa_{\mathcal{I}_E}(t):=
\int_{\mathcal{O}_E^\pm}\idty_{\{\tau_{\mathcal{I}_E}\geq t\}}\Omega_E
\qmbox{of the sets}
V_{\mathcal{I}_E}^\pm(t):=\left\{x\in \mathcal{O}_E^\pm:\tau_{\mathcal{I}_E}(x)\geq t\right\}.
\label{eq:kappa}
\eeq
The motivation for studying this function is that it is 
controllable by symbolic dynamics, and at the same time it is 
asymptotically near to the function 
$\kappa_{E}^\infty$ (defined in (\ref{eq:time_delay_as})), 
which gives the escape rate:
\begin{lem}\label{lem:infty:empty}
$\kappa_{E}^\infty\asymp \kappa_{\mathcal{I}_E}$.
\end{lem}
\pr The $\Phi_t$-invariance of $\tau_{\mathcal{I}_E}$ permits us to define the
\emph{asymptotic total inner Poincar\'e time}
\begin{equation*}
\tau_{\mathcal{I}_E}^+:T^*S^2\rightarrow [0,\infty]\qmbox{,}
\tau_{\mathcal{I}_E}^{+}(\hat{p},\vec{q\,}_\perp):=\left\{\begin{array}{ccc}
\tau_{\mathcal{I}_E}\circ \big(A_E^{+}\big)^{-1}(\hat{p},\vec{q\,}_{\perp})&
\mbox{if } (\hat{p},\vec{q\,}_{\perp})\in D_E^{+} \\
\infty & \mbox{else} \end{array}\right..
\end{equation*}
By Proposition 9.2 of \cite{knauf} and Theorem 10.6 of \cite{klein&knauf} (adapted to
three dimensions) it follows that there exists an energy dependent
constant $C(E)>0$ such that
$|\tau_E(x)-\tau_{\mathcal{I}_E}(x)|\leq C(E)$ uniformly for all
$x\in s_E$. So (setting $\infty-\infty=0$) we have the uniform estimate
$$|\tau^+_E(x)-\tau^+_{\mathcal{I}_E}(x)|<C(E)\qquad(x\in T^*S^2).$$
Thus for the $\Omega_E$-volume of the set
$V_{\mathcal{I}_E}^{\infty}(t):=\{x\in T^*S^2:\tau^+_{\mathcal{I}_E}(x)\geq t\}$ it holds
\beq\label{eq:kappa:inf}
\kappa_{E}^\infty(t)\asymp\int\limits_{T^*
S^2}\idty_{\{\tau^+_{\mathcal{I}_E}\geq t\}}\Omega_E.
\eeq

By Remark \ref{rem:asymptotic_completeness} the set of points $x\in T^*S^2$ with 
$\tau_{\mathcal{I}_E}^+(x)=\infty$ has measure zero.

The asymptotic maps $A_E^\pm$ map the outer Poincar\'e surfaces $\mathcal{O}_E^\pm$
diffeomorphically to their images $A_E^\pm(\mathcal{O}_E^\pm)\subset T^*S^2$.
By using the cotangential lift of the polar diffeomorphism
$\mathbb{R}^3\setminus\{0\}\rightarrow [0,\infty)\times S^2$, see Section 6.3 of
Marsden and Ratiu \cite{marsden},
one can compute that for the pullback with $A_E^\pm$ of the volume forms
(which are derived from the symplectic forms) holds
$({A_E^\pm})^*\Omega_E\rstr_{\mathcal{O}_E^\pm}
= \Omega_E\rstr_{\mathcal{O}_E^\pm}$.
Thus for
the $\Omega_E$-volume $\kappa_{\mathcal{I}_E}(t)$ of the set it
holds for $t>C(E)$, see Remark \ref{rem:bound}
$$\int\limits_{V_{\mathcal{I}_E}^{\infty}(t)}\Omega_E=
\int\limits_{A_E^\pm(V_{\mathcal{I}_E}^\pm(t))}\!\!\Omega
=\int\limits_{V_{\mathcal{I}_E}^\pm(t)}(A_E^\pm)^*\Omega
=\int\limits_{V_{\mathcal{I}_E}^\pm(t)} \Omega_E
=\;\kappa_{\mathcal{I}_E}(t).$$

With Eq.\ (\ref{eq:kappa:inf}) it follows finally that
$\kappa_{E}^\infty\asymp\kappa_{\mathcal{I}_E}$. \eop

In the following we show the estimate $\kappa_{\mathcal{I}_E}(t)\asymp e^{-\er t}$
for the $\Omega_E$-volume of the sets  $V^{\pm}_{\mathcal{I}_E}(t)\subset\mathcal{O}_E^\pm$.
As a first step we define for time $t>0$ the sets of {\em best fitting
words}
\label{pg:best_fitting_sequences}
\begin{eqnarray*} 
\underline{X}_{t,E}&:=&
\{(k_0,\ldots, k_m)\in X^*:
 \forall\ \underline{l}\in [\underline{k}]:\ S_m T_E(\underline{l})\geq t
\qmbox{and}  \exists\ \underline{l}\in [\underline{k}] :S_{m-1} T_E(\underline{k})< t\}
\end{eqnarray*}
and
\begin{eqnarray*}
\overline{X}_{t,E}&:=&
\{(k_0,\ldots, k_m)\in X^*:
\exists\ \underline{l}\in [\underline{k}] :\ S_{m} T_E(\underline{k})\geq t
\qmbox{and}
\forall\  \underline{l}\in [\underline{k}]:\ S_{m-1} T_E(\underline{l})<t\}
\end{eqnarray*}
as subsets of words in $X^*$, with the summatory function of 
$f:X\rightarrow\mathbb{C}$
$$S_0 f:=0\qmbox{,}
S_m f:=\sum_{i=0}^{m-1} f\circ \sigma^i \qquad (m\in\bN). $$
\begin{lem}\label{lem:partition}
 The sets of cylinders
 $[\underline{X}_{t,E}]:=\{[\underline{k}]: \underline{k}\in \underline{X}_{t,E}\}$
 respectively
 $[\overline{X}_{t,E}]:=\{[\underline{k}]: \underline{k}\in \overline{X}_{t,E}\}$
 constitute partitions of $X$.
\end{lem}
\pr
$\bullet$
To show that $[\underline{X}_{t,E}]$ covers $X$, take some $\underline{k}\in X$.
 From Lemma 9.3, Eq.\ (9.21) in \cite{knauf} it follows
$\inf(T_E)>0$. Thus there is a minimal $m\in\mathbb{N}_0$ such that
$S_{m}T_E(\underline{l})\geq t$ for all $\underline{l}\in[(k_0,\ldots, k_m)]$.
Then $(k_0,\ldots, k_m)\in \underline{X}_{t,E}$.\\
$\bullet$
To show that $[\underline{X}_{t,E}]$ is a partition of $X$,
suppose that $\underline{k}=(k_1,\ldots, k_{m(\underline{k})})\in\underline{X}_{t,E}$ and
$\underline{l}=(l_1,\ldots, l_{m(\underline{l})})\in\underline{X}_{t,E}$ with
$[\underline{k}]\cap [\underline{l}]\not=\emptyset$. Then w.l.o.g.\
$[\underline{k}]\subset[\underline{l}]$, i.e.\ $m(\underline{k})\geq m(\underline{l})$.
For all $\underline{x}\in[\underline{l}]$ it holds
$S_{m(\underline{l})}T_E(\underline{x})\geq t$, and there exists an
$\underline{x}\in[\underline{k}]\subset [\underline{l}]$ with
$S_{m(\underline{k})-1}T_E(\underline{x})< t$. Together this implies that
$m(\underline{l})\geq m(\underline{k})$. So $m(\underline{l})= m(\underline{k})$ and
$\underline{k}=\underline{l}$.\\
$\bullet$
The proof for  $[\overline{X}_{t,E}]$ is analogous. \eop

With the aid of the 'best fitting words' we approximate $V^-_{\mathcal{I}_E}(t)$, defined in (\ref{eq:kappa}),
in the following manner:
\begin{prop}\label{prop:channels}
There exists a constant $C_{\tau}(E)>0$  such that for any $t>0$ the inclusions
\beq
 \dot{\bigcup_{\underline{k}\in \underline{X}_{t+C_\tau,E}}}\left(\mathcal{O}_E^{-}
 \cap \mathcal{P}_E^ {-1}(\mathcal{I}_E(\underline{k}))\right)
 \subset V_{\mathcal{I}_E}^-(t)\subset
 \dot{\bigcup_{\underline{k}\in \overline{X}_{t-C_\tau,E}}}\left(\mathcal{O}_E^{-}
 \cap\mathcal{P}_E^{-1}(\mathcal{I}_E(\underline{k}))\right)
\label{inclusions}
\eeq
hold for the iterated inner Poincar\'{e} surfaces $\mathcal{I}_E(\underline{k})$,
defined in (\ref{I:E}).
\end{prop}
\pr
$\bullet$
By the hyperbolicity of $\mathcal{P}_E$ on $\mathcal{I}_E\cap b_E$,
see \cite{knauf}, and the Lipschitz-continuity of $T_E$ on $\mathcal{I}_E$
it follows that there exists a positive constant $C_{\tau}=C_{\tau}(E)$ such that
$$\left|\sum_{i=0}^{m-1}T_E(\mathcal{P}_E^i(x))-\sum_{i=0}^{m-1}T_E(\mathcal{P}^i_E(y))\right|
  < C_{\tau}$$
for all words $\underline{k}\in X^*$ and 
any $x,y\in\mathcal{I}_E(\underline{k})$, 
$m+1$ being the length of $\underline{k}$.\\
$\bullet$
To show the first inclusion in (\ref{inclusions}), take some
$x\in \mathcal{P}_E^{-1}(\mathcal{I}_E(\underline{k}))\cap \mathcal{O}_E^-$
for some $\underline{k}=(k_0,\ldots, k_{m(\underline{k})})\in \underline{X}_{t+C_{\tau},E}$.
 From $\sum\limits_{i=0}^{m-1}T_E(\mathcal{P}_E^i(y))\geq t+C_{\tau}$ for all
$y\in \mathcal{I}_E(\underline{k})\cap b_E$ it follows that
$\sum\limits_{i=0}^{m-1}T_E(\mathcal{P}_E^i(y))\geq t$ for all
$y\in \mathcal{I}_E(\underline{k})$. $\mathcal{P}_E(x)\in \mathcal{I}_E(\underline{k})$
together with $\tau_{\mathcal{I}_E}(x)=\tau_{\mathcal{I}_E}(\mathcal{P}_E(x))$
imply that $x\in V_E^-(t)$. This shows the first inclusion.\\
$\bullet$
To show the second inclusion in (\ref{inclusions}), suppose that $t>0$  and let $x\in
V_{\mathcal{I}_E}^-(t)$. Then there exists $m\in\mathbb{N}$ with
$\sum_{i=0}^{m-1}T_E(\mathcal{P}^{i}_E(\mathcal{P}_E(x)))\geq t$ and
a word $\underline{k}=(k_0,\ldots, k_m)\in X^*$ such that
$\mathcal{P}_E^{i+1}(x)\in \mathcal{I}_E(k_i)$ for $i=0,\ldots,m$.
Since $\mathcal{P}_E(x)\in \mathcal{I}_E(k_0,\ldots, k_m)$, it
follows that there exists
$y\in\mathcal{I}_E(k_0,\ldots,k_m)\cap b_E$ such that
$\sum_{i=0}^{m-1}T_E(\mathcal{P}^{i}_E(y))\geq t-C_{\tau}$. So
after eventually shortening the finite sequence $(k_0,\ldots,
k_m)\in X^*$ to length $m'\leq m$ there exists
$\underline{k}'=(k_0,\ldots, k_{m'})\in
\overline{X}_{t-C_{\tau},E}$ such that $\mathcal{P}_E(x)\in
\mathcal{I}_E(\underline{k})\subset\mathcal{I}_E(\underline{k}') $
and so $x\in\mathcal{P}_E^{-1}(\mathcal{I}_E(\underline{k}'))\cap
\mathcal{O}_E^-$ for some $\underline{k}'\in
\overline{X}_{t-C_{\tau},E}$ showing the second inclusion.\\
$\bullet$
The fact that these unions are disjoint follows from Lemma \ref{lem:partition}
and completes the proof.\hfill$\Box$
\subsubsection{Measure-Theoretical Estimates}
%
%
In order to estimate $V_{\mathcal{I}_E}^-(t)$ using
(\ref{inclusions}),
we now approximate the $\Omega_E$-volume of
$\mathcal{O}_E^-\cap \mathcal{P}_E^{-1}(\mathcal{I}_E(\underline{k}))$
for the words $\underline{k}\in X^*$.

In \cite{knauf} local coordinates $(\y,\z)=(y_1,y_2,z_1,z_2)$
on the inner Poincar\'{e} surfaces $\mathcal{I}^{i,j}_E$ were introduced, 
with $\z$ affine in the position $\q$, $\y$ affine in the momentum $\p$, 
and the volume form
\beq
dy_1\wedge dy_2\wedge dz_1\wedge dz_2 
= \frac{2d_{i,j}^2}{E}\;\Omega_E\rstr_{\mathcal{I}_E^{i,j}} .
\Leq{vol:vol}
By using the Euclidean metric on $\bR^4$, these coordinates serve also for 
defining a metric $d_{\mathcal{I}_E}$ on $\mathcal{I}_E$.
In these coordinates the Poincar\'{e} surfaces defined in
(\ref{eq:surfaces}) take the form
$\mathcal{I}^{i,j}_E=B_y\times B_z$, $B_y$ and $B_z$ being two-dimensional disks
whose radii are proportional to $1/E$.

With $f$ from (\ref{def:f}),
the linearized Poincar\'{e} map equals
\beq
T_x\Po = f(k_0,k_1)\, E\, \bsm
\idty&\idty\\\idty&\idty\esm\ +\ \cO(E^0)\qquad (x\in \mathcal{I}_E(k_0,k_1)), 
\Leq{Tx}
see Prop.\ 11.2 of \cite{knauf}.
In order to use symbolic dynamics, we compare these maps with 
the ones along the bounded orbits. 
Since $\mathcal{I}_E\cap b_E$ is a hyperbolic set for $\mathcal{P}_E$,
the tangent space has the $T\Po$-invariant splitting
\beq
T_x\mathcal{I}_E=T_x^u\oplus T_x^s
\qquad(x\in\mathcal{I}_E\cap b_E)
\Leq{su:bundle}
into the unstable and stable Lagrangian subspace.

In $(\y,\z)$--coordinates, the {\em cone field} $\mathcal{C}^+_E\equiv\cC_E$ 
on $\mathcal{I}_E$ is defined by
\beq
\cC_E(x) :=
\Big\{(\delta\y,\delta\z)\in T_{(\y,\z)} \mathcal{I}_E \; : \;
|\delta\y-\delta\z|\leq \frac{C}{E}\, |\delta\y+\delta\z|\;\Big\},
\Leq{cone}
and $\mathcal{C}_E^-$ denotes the image of $\mathcal{C}^+_E$ under time 
reversal.
So their aperture is of order $\mathcal{O}(1/E)$.

Furthermore, by estimate (\ref{Tx}), for an appropriately chosen
constant $C>0$, for all $E>\Eth$
the cone field (\ref{cone}) is strictly $\mathcal{P}_E$--invariant
on $\mathcal{I}_E\cap \mathcal{P}_E^{-1}(\mathcal{I}_E)$.

The {\em logarithmic Jacobian} of a subbundle $U$
of $T\left(\mathcal{I}_E\cap\mathcal{P}_E^{-1}(\mathcal{I}_E)\right)$
is generally defined by
\beq
F_{E,U}:\mathcal{I}_E\cap\mathcal{P}_E^{-1}(\mathcal{I}_E)\rightarrow \mathbb{R}\qmbox{,}
F_{E,U}(x)=\ln\left(\det\left(T_x\mathcal{P}_E\rstr_{U_x}\right)\right),
\label{FE:u}
\eeq
with the determinant depending on the choice of the Riemannian metric $d_{\mathcal{I}_E}$.
For the two-dimensional Lagrangian unstable
bundle in (\ref{su:bundle}) the {\em logarithmic unstable Jacobian} is
given by
\beq
F_E:\mathcal{I}_E\cap b_E\rightarrow \mathbb{R}^+\qmbox{,}
F_E:=F_{E,T^u}.
\label{FE}
\eeq

Let $\tilde{U},\tilde{V}$ be two transversal Lagrangian subbundles on $\mathcal{I}_E$.
Then any tangent vector $w\in T_x\mathcal{I}_E$ has a unique decomposition $w=u+v$
with $u\in \tilde{U}$ and $v\in \tilde{V}$.\\
This decomposition gives rise to the quadratic form $Q_E(w):=2\omega(u,v)$ and
defines the sector field
$$\mathcal{S}_E:=
\bigcup\limits_{x\in\mathcal{I}_E}\{w\in T_x\mathcal{I}_E: Q_{E}(w)\geq 0\},$$
both depending on the pair $\tilde{U},\tilde{V}$.

There exist two Lagrangian subbundles $\tilde{U},\tilde{V}$
such that for their sector it holds: $\mathcal{C}_E\subset \mathcal{S}_E$
(for example, one can choose $\tilde{V}$ to be tangential to the kernel of the cotangent
bundle projection $T^*\hat{M}\rightarrow \hat{M}$ and $\tilde{U}$ to be horizontal w.r.t.\
the Euclidean metric, see Sect.\ 11 of \cite{knauf}).
In the following we denote with $\mathcal{S}_E$ the sector defined by such
a pair of transversal Lagrangian subbundles.

\begin{lem}
 For any $E>\Eth$ the tangent map of the Poincar\'e map $\mathcal{P}_E$ is
 \emph{strictly monotone with respect to the sector $\mathcal{S}_E$}, i.e.\
 $T_x \mathcal{P}_E (\mathcal{S}_{E,x}\setminus\{0\})\subset \mbox{int} \mathcal{S}_{E,\mathcal{P}_E(x)}$
 for any $x\in \mathcal{I}_E\cap \mathcal{P}_E^{-1}(\mathcal{I}_E)$.
\end{lem}
\pr 
Let $v\in \mathcal{S}_{E,x}\setminus \{0\}$ for some point
$x\in\mathcal{I}_E\cap \mathcal{P}_E^{-1}(\mathcal{I}_E)$.
Since $T^u_y\mathcal{I}_E\subset \mathcal{C}_{E,y}\subset \mathcal{S}_{E,y}$
and the logarithmic unstable Jacobian of $T_y\mathcal{P}_E$ is of order $E$ for
$y\in\mathcal{I}_E\cap b_E$ the claim follows by a compactum argument.
\eop\\[2mm]
Following Liverani and Wojtkowski \cite{liverani}, we denote for $x\in\mathcal{I}_E$ by
$$\mathcal{L}_{E,x}:=\{ \mbox{Lagrangian subspace } U\subset T_x\mathcal{I}_E:
\forall u\in U\setminus\{0\}:\  Q_{E,x}(u)>0\}$$
the set of \emph{positive} Lagrangian subspaces. With $\mathcal{L}_{E}$ we denote the set of
positive Lagrangian subbundles of $T\mathcal{I}_E$.
For two subspaces $U,V\in\mathcal{L}_{E,x}$ the distance
$$s_{E,x}(U,V):=\sup_{
{u\in U\setminus\{0\},\
v\in V\setminus\{0\}}}\
\left|\mbox{Asinh}\left(\frac{\omega_x(u,v)}{\sqrt{Q_{E,x}(v)}\sqrt{Q_{E,x}(w)}}\right)\right|$$
gives a complete metric on $\mathcal{L}_{E,x}$, see \cite{liverani}.

By Theorem 1 of \cite{liverani} and the strict monotonicity of the Poincar\'e map
$\mathcal{P}_E$ it follows that $Q_{E}(T_x\mathcal{P}_E(v))> Q_{E}(v)$ for any
$v\in \mathcal{C}_{E,x}$, $v\not=0$ and thus
$$\gamma_E:= \sup_{x\in {\mathcal{I}_E}\cap\mathcal{P}_E^{-1}(\mathcal{I}_E)}\
\sup_{v\in \mathcal{S}_{E,x}:\|v\|_{\mathcal{I}_E}=1}\frac{Q_E(v)}{Q_E(T_x\mathcal{P}_E(v))}<1$$
by a compactness argument. Thus for any $u,v\in \mbox{int}{\mathcal S}_{E,x}$ it holds
(note that $\mathcal{P}_E$ is a symplectomorphism)
that
$$\left|\mbox{Asinh}\left(
   \frac{\omega(T_x\mathcal{P}^m_E(u),T_x\mathcal{P}_E^m(v))}
	{\sqrt{Q_E(T_x\mathcal{P}^m_E(u))Q_E(T_x\mathcal{P}^m_E(v))}}\right)\right|
\leq \left|\mbox{Asinh}\left(
   \frac{\omega(u,v)}{\sqrt{Q(u) Q(v)}}\gamma_E^m\right)\right|
   \stackrel{m\rightarrow\infty}{\longrightarrow} 0$$ implying
$\lim_{m\rightarrow\infty}s_{E,\mathcal{P}_E^m(x)}(T_{x}\mathcal{P}_E^m(U), T_{x}\mathcal{P}_E^m(V))=0$
for any positive Lagrangian subspaces $U,V\in\mathcal{L}_{E,x}\subset T_x\mathcal{I}_E$ and $x\in\mathcal{I}_E\cap b_E$.

Since the set of positive Lagrangian subspaces lying in the cone $\mathcal{C}_{E,x}$ is compact,
it follows that there exists a constant $C_{s,E}>0$ such that
\beq\label{eq:C_sE}
 s \big(T_x \mathcal{P}_E^m(U),T_x\mathcal{P}_E^m(V)\big)
 \leq C_{s,E}\cdot\gamma_E^m\qquad
 \left(x\in\cap_{i=0}^{m}\mathcal{P}_E^{-i}(\mathcal{I}_E)\right).
\eeq

Given a two-dimensional subspace $U_x\subset T_x\mathcal{I}_E$ we denote with
$F_{E,U_x}(x)$ the logarithm of the Jacobian of $T_x\mathcal{P}_E$ restricted to $U_x$
with  respect to the metric $d_{\mathcal{I}_E}$.\\
For a two-dimensional subbundle $U$ of $T\mathcal{I}_E$ and $m\geq 1$ we
have
\beq
S_m F_{E,U}(x):=\sum_{i=0}^{m-1} F_{E,T_x\mathcal{P}_E^i(U_x)}\big(\mathcal{P}_E^i(x)\big)
\qquad
\left(x\in\cap_{i=0}^{m-1}\mathcal{P}_E^{-i}(\mathcal{I}_E)\right).
\Leq{S:m}
Note that $S_m F_{E,U}(x)$
is the logarithm of the Jacobian of $\mathcal{P}_E^m$ restricted to
$U_x\subset T_x\mathcal{I}_E$, $x\in\cap_{i=0}^{m}\mathcal{P}_E^{-i}(\mathcal{I}_E)$.

In order to estimate the volumes in (\ref{inclusions}), in the proof
of Prop.\ \ref{prop:size} we control $S_m F_{E,U}$ for a concrete
positive Lagrangian subbundle $U\in \mathcal{L}_E$.
The following lemma relates this to
$S_m F_E$, defined on the much smaller set $\mathcal{I}_E\cap b_E$.
\begin{lem}\label{lem:C_FE}
 There exists a constant $C_{F_E}$ such that for any $x,y\in\mathcal{I}_E(\underline{k})$,
 $x\in b_E$ with $\underline{k}=(k_0,\ldots, k_m)\in X^*$ and any smooth positive Lagrangian
 subbundle $V\in \mathcal{L}_{E}\subset \mathcal{C}_E$ it holds:
$$\left|S_m F_{E,V}(y)-S_m F_E(x)\right|<C_{F_E}.$$
\end{lem}
\pr 
Let $x,y\in\mathcal{I}_E(\underline{k})$, $x\in b_E$ with $\underline{k}=(k_0,\ldots, k_m)\in X^*$. Then we have
\beq\label{eq:sum}
|S_m F_E(x)-S_m F_{E,V}(y)|\leq |S_m F_E(x)-S_m F_{E,V}(x)| + |S_m F_{E,V}(x)-S_m F_{E,V}(y)|.
\eeq

Since for any $x\in\mathcal{I}_E\cap
\mathcal{P}_E^{-1}(\mathcal{I}_E)$ the set $(\mathcal{L}_{E,x},s_E)$
of positive Lagrangian subspaces in the cone $\mathcal{C}_{E,x}$ is a
compact metric space and since the topology defined by this metric
coincides with the standard topology, see Corollary on p.\ 8 of \cite{liverani}, the map
$\mathcal{L}_{E,x}\rightarrow \mathbb{R}$, $U\mapsto F_{E,U}(x)$ is continuously differentiable.
Thus  it holds by a compactness argument that this map is Lipschitz continuous, i.e.\
\begin{equation*}
 |F_{E,U}(x)-F_{E,V}(x)|\leq C\cdot s_E(U_x,V_x)
\end{equation*}
for an appropriate constant $C>0$. Hence by Eq.\ (\ref{eq:C_sE})
$$|S_m F_E(x)-S_m F_{E,V}(x)|
\leq\sum_{i=0}^{m-1}
\left| F_{E,\mathcal{P}_E^i(T^u_x\mathcal{I}_E)}\big(\mathcal{P}_E^i(x)\big)
-F_{E,T\mathcal{P}_E^i(V)}\big(\mathcal{P}_E^i(x)\big) \right| $$
$$\leq C\sum_{i=0}^{m-1} s_E\big(T_x\mathcal{P}_E^i(V_x),T_x\mathcal{P}_E^i(T^u_x\mathcal{I}_E)\big)
\leq C\cdot C_{s,E}\sum_{i=0}^{m-1}\gamma_E^i\leq \frac{C\cdot C_{s,E}}{1-\gamma_E},$$
showing that the first summand in Eq.\ (\ref{eq:sum}) is bounded uniformly in $m$.

The second sum in Eq.\ (\ref{eq:sum}) is bounded by a constant by the
fact that the function $x\mapsto F_{E,U_{x}}$ is smooth on
$\mathcal{I}_E\cap \mathcal{P}_E^{-1}(\mathcal{I}_E)$, and thus
also Lipschitz-continuous on
$\overline{\mathcal{I}_E\cap \mathcal{P}_E^{-1}(\mathcal{I}_E)}$ by a compactness argument. \eop

Note that by the chain rule of differentials
$S_m F_E(x)=\sum_{i=0}^{m-1} F_E(\mathcal{P}_E^i(x))$ is the
logarithm of the unstable Jacobian of $\mathcal{P}_E^m(x)$ for
$x\in\mathcal{I}_E\cap b_E$.
\begin{prop}\label{prop:size}
With $F_E:\mathcal{I}_E\cap b_E\rightarrow \mathbb{R}^+$ the logarithm of
the unstable Jacobian it holds: 
There exists a constant $C>1$ such that for all $E>\Eth$
\beq
C^{-1}\; \frac{\exp(-S_m F_E(x))}{E^3}
\leq \Omega_E\big(\mathcal{I}_E(\underline{k})\big)
\leq C\;  \frac{\exp(-S_m F_E(x))}{E^3}
\Leq{eq:size_jacobian_shift_space}
uniformly for any $\underline{k}=(k_0,\ldots, k_m)\in X^*$ and  $x\in \mathcal{I}_E(\underline{k})\cap b_E$.
\end{prop}
\pr 
$\bullet$
For the word $\underline{k}\in X^*$ and $\ell=0,\ldots,m$
we denote by $(\vec{y}_\ell,\vec{z_\ell})$
the local $(\vec{y},\vec{z})$--coordinates on $\mathcal{I}_E(k_\ell)$. 
By (\ref{vol:vol}) the canonical volume form $\Omega_E$ appearing in 
(\ref{eq:size_jacobian_shift_space})
and the standard coordinate area forms 
$\Omega_{y} := dy_1\wedge dy_2$ on the disk $B_y$, 
$\Omega_{z} := dz_1\wedge dz_2$ on $B_z$
are related by
$\Omega_E\rstr_{\mathcal{I}_E(k_0)}=\frac{E}{2d_{k_0}^2}
\Omega_{{y}_0}\wedge \Omega_{{z}_0}$.

For all points $\vec{z}_0\in B_z$ the set
$B_{y_0}(\vec{z}_0,\underline{k})
:=\{x=(\vec{y},\vec{z})\in\mathcal{I}_E(\underline{k}):\vec{z}=\vec{z}_0\}$ 
is a two-disk, and is identified with its image in $B_y$.
Then
\beq
\int_{\mathcal{I}_E(\underline{k})}\Omega_E
=\frac{E}{2d_{k_0}^2}\int_{B_{{z}}}\left(\int_{B_{y_0}(\vec{z}_0,\underline{k})}\Omega_{{y}_0}\right) \Omega_{{z}_0}.
\Leq{eq:omega_integral}
We restrict the iterated Poincar\'{e} maps $\mathcal{P}_E^m$ to the
two-disks $B_{y_0}(\vec{z}_0,\underline{k})\subset \mathcal{I}_E(k_0)\quad(\vec{z}_0\in B_{z_0}$), and 
denote by 
$\pi_{\vec{z}_0}:
\mathcal{P}_E^m\big(B_{y_0}(\vec{z}_0,\underline{k})\big)\rightarrow B_{z_m}$ 
the projections 
of their images to the ${z}_m$--coordinate plane.
By Prop.\ 11.5 (1) of \cite{knauf} the composition of these maps
gives rise to the diffeomorphisms 
\beq
\mathcal{Z}_{\vec{z}_0,\underline{k}}:=
\pi_{\vec{z}_0}\circ\mathcal{P}_E^m\rstr_{B_{y_0}(\vec{z}_0,\underline{k})}
:B_{y_0}(\vec{z}_0,\underline{k})
\ \rightarrow \ B_{z_m} \qquad(\vec{z}_0\in B_{z_0}).
\Leq{Z:diffeo} 
In the inner integral on the right hand side of 
(\ref{eq:omega_integral}) we apply 
the transformation rule of integration
\beq
\int_{B_{{y}_0}(\vec{z}_0,\underline{k})}\Omega_{{y}_0}
=\int_{\mathcal{Z}_{\vec{z}_0,\underline{k}}^{-1}(B_{{z}_m})}\Omega_{{y}_0}
=\int_{B_{{z}_m}}{\mathcal{Z}_{\vec{z}_0,\underline{k}}\,}_*\,\Omega_{{y}_0}
\Leq{eq:integration_rule}
$\bullet$
We denote by $\mathcal{V}_E$ the {\em vertical bundle} whose
form in $(\y,\z)$--coordinates is
\[\mathcal{V}_{E,x}:= \l\{ (\delta\y,\delta\z) \in T_x\mathcal{I}_E\; :\;
\delta\z=0\ri\}\qquad
\l(x\in\mathcal{I}_E\cap\mathcal{P}_E^{-1}(\mathcal{I}_E)\ri).\]

Using the notation (\ref{S:m}), the push-forward of the two-form 
$\Omega_{{y}_0}$ with the 
diffeomorphism (\ref{Z:diffeo}) equals for 
$x\in B_{y_0}(\vec{z}_0,\underline{k})$ 
\beq
\big( {\mathcal{Z}_{\vec{z}_0,\underline{k}}\,}_*\;
\Omega_{{y}_0}\big)\big(\mathcal{Z}_{\vec{z}_0,\underline{k}}(x)\big)
=\exp\big(-S_m F_{E,\mathcal{V}_{E}}(x)\big)\,
\big(J_{\pi_{\vec{z}_0}}\mathcal{P}_E^m(x)\big)^{-1}\,
\Omega_{{z}_m}\big(\mathcal{Z}_{\vec{z}_0,\underline{k}}(x)\big),
\Leq{three:factors}
with $J_{\pi_{\vec{z}_0}}$ the Jacobian of the projection 
$\pi_{\vec{z}_0}$,
of the surface $S:=\mathcal{P}_E^m(B_{y_0}(\vec{z}_0,\underline{k}))$.
We insert (\ref{three:factors}) in (\ref{eq:integration_rule}),
estimating its constituents.
\\
$\bullet$
That Jacobian in (\ref{three:factors}) is estimated uniformly in 
the parameter $\vec{z}_0\in B_{z_0}$ by 
\beq
J_{\pi_{\vec{z}_0}}=\eh+\cO(1/E).
\Leq{est:jac}
To show this we choose an orthonormal basis $w^{(1)},w^{(2)}$
of $T_aS$ at $a:=\mathcal{P}_E^m(x)$. Then the tangent vectors are
contained in the local cone at $a$. 
Writing 
$w^{(1)}=(\delta y^{(i)},\delta z^{(i)})$, this implies
that $\l| |\delta y^{(i)}|^2-\eh\ri|=\l| |\delta z^{(i)}|^2-\eh\ri|\le C/E$,
with $C$ from (\ref{cone}).
On the other hand 
\[\LA \delta y^{(1)},\delta y^{(2)}\RA
=-\LA \delta z^{(1)},\delta z^{(2)}\RA\]
by orthogonality of $w^{(1)}$ and $w^{(2)}$. With (\ref{cone})
this implies
$\l| \LA \delta z^{(1)},\delta z^{(2)}\RA\ri|=\cO(1/E)$.
As $J_{\pi_{\vec{z}_0}}
=\sqrt{|\delta z^{(1)}|^2\,|\delta z^{(2)}|^2-\LA \delta z^{(1)},\delta
z^{(2)}\RA^2}$, we have proven
(\ref{est:jac}).\\
$\bullet$
Note that both the unstable bundle
$T^{u}(\mathcal{I}_E\cap b_E)$ and the iterates of the vertical bundle $\mathcal{V}_E$
are contained in the cone $\mathcal{C}_E$. 
Furthermore the smoothness of the map $x\mapsto F_{E,\mathcal{V}_E}(x)-F_{E,V}$ 
for $V\in\mathcal{L}_E$ a positive Lagrangian subbundle together with compactum
argument and Lemma \ref{lem:C_FE} shows the existence of a constant 
$C$ such that the first factor on the right hand side of 
(\ref{three:factors}) is estimated by
\beq
C^{-1} \exp\big(-S_m F_E(x_0)\big)
\leq \exp\big(-S_m F_{E,\mathcal{V}_{E}}(x)\big)
\leq C \exp\big(-S_m F_E(x_0)\big)
\Leq{sss}
uniformly in $E>\Eth$ and for any $x\in\mathcal{I}_E(\underline{k})$, 
$x_0\in\mathcal{I}_E(\underline{k})\cap b_E$. \\
$\bullet$
Insertion of (\ref{est:jac}) and (\ref{sss}) in 
(\ref{three:factors}) completes  the proof,
taking into regard the fact that $\Omega_{{y}_0}(B_y)\asymp 1/E^2$ 
and $\Omega_{{z}_m}(B_z)\asymp 1/E^2$ . 
\eop

Propositions \ref{prop:channels}
respectively  \ref{prop:size} concern subsets of the outer resp.\
inner Poincar\'{e} surfaces. To combine them  
note that the sets appearing in Prop.\ \ref{prop:channels} can be
written as
\beq
\mathcal{O}_E^-\cap\mathcal{P}_E^{-1}\big(\mathcal{I}_E(\underline{k})\big)
 =\mathcal{P}_E^{-1}\big(\mathcal{I}_E(\underline{k})\big)\setminus
  \mathcal{I}_E.
\Leq{double} 
Since $\Omega_E$ is $\mathcal{P}_E$-invariant, 
this trivially implies the upper bound in
\beq
\eh\Omega_E\big(\mathcal{I}_E(\underline{k})\big)
\leq \Omega_E\Big(\mathcal{O}_E^-\cap \mathcal{P}_E^{-1}\big(\mathcal{I}_E(\underline{k})\big)\Big)
\leq \Omega_E\big(\mathcal{I}_E(\underline{k})\big)\qquad 
(E>\Eth,\underline{k}\in X^*).
\Leq{upper:lower}
The volume of the r.h.s.\ in (\ref{upper:lower})
equals 
$\Omega_E\big(\mathcal{I}_E(\underline{k})\big)- 
\Omega_E\big(\mathcal{P}_E^{-1}\big(\mathcal{I}_E(\underline{k})\big)\cap
  \mathcal{I}_E\big)$.

The explicit formula (\ref{Tx})
for the differential $T\mathcal{P}_E$
shows that the unstable Jacobian diverges (like $E^2$) 
as $E\rightarrow\infty$
and so does its logarithm $F_E$.

So the lower bound in (\ref{upper:lower})
(even with any constant smaller than one instead of 
$1/2$) also follows 
for large enough threshold energy $\Eth$
from (\ref{double}) and the estimate of Proposition \ref{prop:size} in terms of scaling 
factors $F_E$.

Thus for an appropriate constant $C>1$ we obtain the estimate
\beq
 C^{-1}\sum_{\underline{k}\in \underline{X}_{t+C_\tau,E}}
 \exp\big(-S_{m(\underline{k})}F_E (\underline{x}_{\underline{k}})\big)
 \leq \kappa_{\mathcal{I}_E}(t)
 \leq C\sum_{\underline{k}\in \overline{X}_{t-C_{\tau},E}}
 \exp\big(-S_{m(\underline{k})}F_E (\underline{x}_{\underline{k}})\big)
\Leq{eq:estimate_kappa_IE_bestfitting}
valid for all $E>\Eth$ and arbitrary representatives  
$\underline{x}_{\underline{k}}\in [\underline{k}]\subset X$.

The hyperbolicity of $b_E$ assures that also the logarithm of the
unstable Jacobian $F_E$ is H\"older continuous on $\mathcal{I}_E\cap
b_E$ resp.\ on $X$, see Thm. 19.1.6 and its Corollary 19.1.13 in \cite{katok}.

Like for $T_E$ from (\ref{P:T}) we will,
using the homeomorphism (\ref{symb:homeo}),
consider $F_E: b_E\cap {\cal I}_E\rightarrow \bR^+$ (see (\ref{FE}))
also as an element of
$\mathcal{F}_{\alpha}(X^+,\mathbb{R})$.

Since the following approach depends on the cohomology classes
of the functions $T_E$ and $F_E$ only, we can assume
that $T_E$ and $F_E$ are H\"older
continuous functions on $X$ only depending on the future, 
and by a natural identification
that $T_E$ and $F_E$ are H\"older continuous functions on the
one-sided shift $X^+$. See Bowen \cite{bowen}, Sect.\ 6 for more details.

Next we define, using the one-sided shift $(X^+,\sigma)$,
for $\underline{x}\in X^+$ the function 
\beq
\kappa_{\mathcal{I}_{E}}^{\underline{x}}:\bR\rightarrow\bR^+
\qmbox{,}\kappa_{\mathcal{I}_{E}}^{\underline{x}}
:= \sum_{m=0}^\infty \ \sum_{\underline{y}\in \sigma^{-m}(\underline{x})}
\exp\l({-S_m F_E(\underline{y})}\ri)\ 
\idty_{\l(-\infty,S_m T_E(\underline{y})\ri]}.
\Leq{kappa:x}
This quantity models $\kappa_{\mathcal{I}_E}$, defined in (\ref{eq:kappa}),
but is only based on 
data of bounded orbits.
We start with a rough upper estimate, needed later on for renewal theory.

\begin{lem}\label{lem:renewal_bounds}
For a suitable
energy threshold $\Eth>0$ and $C>0$ for
all energies $E>\Eth$ the sum
$\kappa_{\mathcal{I}_{E}}^{\underline{x}}(t)$ converges for any
$t\in\bR$
and any choice $\underline{x}\in X^+$. Furthermore
$$0<\kappa_{\mathcal{I}_{E}}^{\underline{x}}(t)
\leq C \exp\big(-\omega^-(E) \max(t,0)\big)$$
with
\beq
0 < \omega^-(E) := \frac{\inf (F_E)-\htop(X^+,\sigma)}{\sup(T_E)}\ \le \ 
    \omega^+(E) := \frac{\sup (F_E)-\htop(X^+,\sigma)}{\inf(T_E)},
\Leq{eq:varepsilon}
$\htop(X^+,\sigma)\ge0$ being the ($E$-independent)
topological entropy of the shift space $(X^+,\sigma)$.
\end{lem}
\pr 
$\bullet$
For all $n\ge 2$ there exists a constant $C_h\geq 1$ such that
$$C_h^{-1}\cdot e^{n \htop(X^+,\sigma)}
\leq |\sigma^{-m}(\underline{x})|\leq C_h\cdot e^{m \htop(X^+,\sigma)}
\qquad (n\in\mathbb{N}, \underline{x}\in X^+).$$
For $n=2$ this follows since
the topological entropy $\htop(X^+,\sigma)=0$
and $|\sigma^{-m}(\underline{x})|=1$.
For $n\ge 3$ 
the estimate follows since then the shift
is topological mixing.\\
$\bullet$
We already remarked that
the logarithmic unstable Jacobian $F_E$ diverges
as $E\rightarrow\infty$.
Thus for $\Eth\ge 1$ large enough we have 
\beq
F_E-\htop(X^+,\sigma)\ge\ln(2)\qquad (E>\Eth),
\Leq{logtwo}
which in particular vindicates the first inequality 
in (\ref{eq:varepsilon}).\\ 
$\bullet$
With the constant $C>0$ from Prop.\ \ref{prop:size} and
for $m_0:=\lfloor\max(t,0)/\sup(T_E) \rfloor$
$$\kappa_{\mathcal{I}_E}^{\underline{x}}(t)
\leq \frac{C\,C_h}{E^3} \sum_{m=m_0}^\infty 
e^{m\htop(X^+,\sigma)}e^{-m\inf(F_E)}\le 
\tilde{C}(E)\exp({-\omega^-(E)\max(t,0)})<\infty, $$
with 
$\tilde{C}(E):=2\frac{C\,C_h}{E^3}e^{F_E-\htop(X^+,\sigma)}=\cO(\frac{1}{E})$,
using (\ref{logtwo}) in estimating the geometric series.
\eop

The following lemma shows the asymptotic equivalence of
the functions $\kappa_{\mathcal{I}_E}$ and
$\kappa_{\mathcal{I}_{E}}^{\underline{x}}$:
\begin{lem}\label{lem:kappa_IE_x}
For any energy $E>\Eth$ it holds uniformly in $\underline{x}\in X^+$ that
\beq\label{kappa_IE-x_kappa_IE}
\kappa_{\mathcal{I}_{E}}^{\underline{x}}(t)\asymp \kappa_{\mathcal{I}_{E}}(t).
\eeq
\end{lem}
\pr 
We show the existence of a constant $C>1$ such that for all $E>\Eth$,
and $\underline{x}\in X^+$
\beq
C^{-1}\kappa_{\mathcal{I}_{E}}^{\underline{x}}(t)
\leq E^3\kappa_{\mathcal{I}_{E}}(t)
\leq C\kappa_{\mathcal{I}_{E}}^{\underline{x}}(t)\qquad(t>0).
\Leq{es:es}
We start with the first inequality in (\ref{es:es}).

Since by Lemma \ref{lem:partition}
the cylinders over the set
$\overline{X}_{t+C_{\tau_E},E}\subset [X^+]$ of best fitting words 
constitute a partition of  the shift space $X^+$, it holds:
\beq
\kappa_{\mathcal{I}_E}^{\underline{x}}(t)
=\sum_{\underline{k}\in\overline{X}_{t+C_{\tau_E},E}}\
\Big(\sum_{m=0}^\infty \ \sum_{\underline{y}\in \sigma^{-m}(\underline{x})}
e^{-S_m F_E(\underline{y})}
\idty_{\l(-\infty , S_m T_E(\underline{y})\ri]}(t)\Big)
\cdot\idty_{[\underline{k}]}(y).
\Leq{kiex}
By Eq.\ (\ref{eq:size_jacobian_shift_space}) and the fact that
$S_{m(\underline{k})} {T_E}\rstr_{[\underline{k}]}<t$ it follows
that for an appropriate $C_1>1$ and any best fitting word 
$\underline{k}\in\overline{X}_{t+C_{\tau_E},E}$ 
the corresponding term in (\ref{kiex}) is dominated by
\beqn
\lefteqn{\NN
\sum_{m=0}^\infty\sum_{\underline{y}\in \sigma^{-m}(\underline{x})} 
e^{-S_m F_E(\underline{y})}\idty_{\l(-\infty , S_m T_E(\underline{y})\ri]}(t)
\cdot\idty_{[\underline{k}]}(\underline{y})\leq
\sum_{m=m(\underline{k})}^\infty \
\sum_{\underline{y}\in\sigma^{-m}(\underline{x})} \
e^{-S_m F_E(\underline{y})}\idty_{[\underline{k}]}(\underline{y})}\\
&=&\sum_{m'=0}^\infty\ \sum_{\underline{y}\in
\sigma^{-(m'+m(\underline{k}))}(\underline{x})}
\exp\big[-S_{m(\underline{k})}
F_E(\underline{y})-S_{m'}F_E(\sigma^{m(\underline{k})}(\underline{y}))\big]\idty_{[\underline{k}]}(\underline{y})
\NN\\
&\leq& C_1\;
E^3\;\Omega_E(\mathcal{I}_E(\underline{k}))\;\sum_{m'=0}^\infty\sum_{\underline{y}\in
\sigma^{-m'}(\underline{x})} e^{-m'\inf(F_E)}.\label{series}
\eeqn
Note that $|\sigma^{-m}(\underline{x})|\asymp e^{\htop(X,\sigma)m}$ 
uniformly in $\underline{x}\in X^+$.
Thus since the unstable Jacobian diverges as $E\rightarrow\infty$, the 
double sum in (\ref{series}) converges (if $\Eth$ is chosen large enough) 
for all $E>\Eth$ and $\underline{x}$, with upper bound 2. So the first
estimate in (\ref{es:es}) follows by using (\ref{eq:estimate_kappa_IE_bestfitting}).
 With analogous arguments we get
$$\kappa_{\mathcal{I}_E}^{\underline{x}}(t)\geq
C_1^{-1}E^3\;\Omega_E(\mathcal{I}_E(\underline{k}))\sum_{m'=0}^\infty\sum_{\underline{y}\in
\sigma^{-m'}(\underline{x})} e^{-m'\max\{F_E\}},$$ showing (after
an adaptation on the constant $C$ if necessary) the second estimate
in (\ref{es:es}).
\eop

The key-feature of the function 
$t\mapsto \kappa^{\underline{x}}_{\mathcal{I}_E}(t)$
which allows for a precise study of its asymptotic behaviour is the following:
\begin{lem}
The function $\kappa^{\underline{x}}_{\mathcal{I}_E}:\bR\rightarrow\bR$ 
satisfies the renewal equation
\beq\label{eq:renewal_equation}
\kappa^{\underline{x}}_{\mathcal{I}_E}(t) =\idty_{\{t\leq 0\}}
+\sum_{\underline{z}\in\sigma^{-1}(\underline{x})}e^{-F_E(\underline{z})}
\ \kappa^{\underline{z}}_{\mathcal{I}_E}(t-T_E(\underline{z})).
\eeq
\end{lem}
\pr 
Noting that the sums $S_0T_E=0$, we decompose (\ref{kappa:x}) into
\begin{eqnarray*}
\kappa^{\underline{x}}_{\mathcal{I}_E}(t)&=&\idty_{\{t\leq 0\}}
+\sum_{\ell=1}^\infty \sum_{\underline{y}\in \sigma^{-\ell}(\underline{x})}
\exp(-S_\ell F_E(\underline{y}))\cdot \idty_{\{S_\ell T_E(\underline{y})\geq t\}}\\
&=& \idty_{\{t\leq 0\}}+\sum_{\underline{z}\in\sigma^{-1}(\underline{x})}e^{-F_E(\underline{z})}
\sum_{m=0}^\infty\sum_{\underline{y}\in\sigma^{-m}(\underline{z})}
\exp\l(-S_m F_E(\underline{y})\ri)\cdot \idty_{\{S_{m+1}T_E(\underline{y})\geq t\}}\\
&=& \idty_{\{t\leq 0\}}+\sum_{\underline{z}\in\sigma^{-1}(\underline{x})}
e^{-F_E(\underline{z})} \kappa^{\underline{z}}_{\mathcal{I}_E}(t-T_E(\underline{z})),
\end{eqnarray*}
that is, the renewal equation. 
\eop

For any $E>\Eth$ and $\omega\in\mathbb{C}$ the function $\omega \,T_E - F_E$ 
is H\"older continuous. The associated \emph{Ruelle transfer operator}
\beq
\mathcal{L}_{\omega,E}:\mathcal{C}(X^+,\mathbb{R})\rightarrow \mathcal{C}(X^+,\mathbb{R})
\qmbox{,}
\mathcal{L}_{\omega,E} f(\underline{k}) :=
\sum_{\underline{l}\in\sigma^{-1}(\underline{k})}e^{\omega T_E(\underline{l})-F_E(\underline{l})}f(\underline{l})
\Leq{eq:transfer_operator}
is a Perron-Frobenius (PF) operator if $\omega\in\mathbb{R}$. We denote with 
$\lambda_{\omega,E}$, $h_{\omega,E}$ and $\nu_{\omega,E}$
its PF eigenvalue, 
its normalized positive PF eigenfunction
and its adjoint Borel PF probability measure respectively, i.e.\ 
(omitting the index $E$)
\beq
\mathcal{L}_\omega h_\omega=\lambda_\omega h_\omega\qmbox{,}
\mathcal{L}_\omega^*\nu_\omega=\lambda_\omega \nu_\omega \qmbox{and}
\int_{X^+} h_\omega\, d\nu_\omega =1.
\Leq{normal}
The solution $\omega_0(E)$ 
of the implicit equation $\lambda_{\omega,E}=1$ turns out to 
be the escape rate.
\begin{lem}\label{lem:unique}
For all $E>\Eth$ there exists a unique solution $\omega_0(E)\in\mathbb{R}^+$ of the
equation $\lambda_{\omega,E}=1$, and $\omega_0(E)\in[\omega^-(E),\omega^+(E) ]$
with $\omega^\pm$ from (\ref{eq:varepsilon}).
\end{lem}
\pr 
$\bullet$
The fact that $\omega \mapsto \lambda_\omega$ is continuously differentiable with
\beq\label{eq:diff_lambda}
\frac{d\lambda_\omega}{d\omega}=
\frac{d}{d\omega}\int_X \mathcal{L}_\omega h_\omega\,d\nu_\omega
= \lambda_\omega \int_{X^+} T_E h_\omega\,d\nu_\omega >0,
\eeq
using the normalizations (\ref{normal}), shows uniqueness of the solution.\\
$\bullet$
For existence and localization
first notice that for $\omega\leq \omega^-(E)= \frac{\inf (F_E)-\htop(X^+,\sigma)}{\sup(T_E)}$
and independent of $\underline{k}\in X^+$
\beqno
\lambda_{\omega,E}&=&\lim_{m\rightarrow\infty}
\left(\frac{\mathcal{L}^m_\omega
h_\omega(\underline{k})}{h_\omega(\underline{k})}\right)^{1/m}
=\lim_{m\rightarrow\infty}
\left(\frac{\sum\limits_{\underline{l}\in(\sigma^m)^{-1}(\underline{k})}
\exp\big(S_m(\omega T_E-F_E)
(\underline{l})\big) \ h_\omega(\underline{l})}{h_\omega(\underline{k})}\right)^{1/m}\\
&\leq& \lim_{m\rightarrow\infty}
\left(\frac{C_h\sup(h_\omega)}{\inf(h_\omega)}\
\exp\Big(m\big(\omega-\omega^+(E)\big)\; \sup(T_E)\Big)\right)^{1/m}\le 1,
\eeqno
since for $E>\Eth$ we have $\htop(X^+,\sigma)< F_E$, see (\ref{eq:varepsilon}).\\
$\bullet$
Similarly for 
$\omega\geq \omega^+(E) = \frac{\sup (F_E)-\htop(X^+,\sigma)}{\inf(T_E)}$ 
independent of $\underline{k}\in X^+$
\beqno
\lambda_{\omega,E} &=&\lim_{m\rightarrow\infty}
 \left(\frac{\sum\limits_{\underline{l}\in \sigma^{-m}(\underline{k})}
 \exp\big(S_m(\omega\, T_E-F_E)(\underline{l})\big)\ h_\omega(\underline{l})}
 {h_\omega(\underline{k})}\right)^{1/m}\\
&\geq& \lim_{m\rightarrow\infty}\left(C_h^{-1}
 \frac{\inf(h_\omega)}{\sup(h_\omega)}\
 \exp\Big(m\big(\omega-\omega^+(E)\big)\; \inf(T_E)\Big)\right)^{1/m} 
 \geq 1,
\eeqno
together showing existence of a solution 
$\omega_0(E) \in[\omega^-(E),\omega^+(E)]$. \hfill $\Box$

\begin{prop}
With $\kappa^{\underline{x}}_{\mathcal{I}_E}$ from
(\ref{kappa:x}) and $\omega_0(E)$ from Lemma \ref{lem:unique} there exists a 
constant $C\geq 1$ such that
$$C^{-1} K_E(\underline{x})\;\exp\big(-\omega_0(E) t\big)
\leq \kappa^{\underline{x}}_{\mathcal{I}_E}(t)
\leq C K_E(\underline{x})\;\exp\big(-\omega_0(E) t\big)
\qquad (t\in\bR^+,\,\underline{x}\in X^+),$$
$$K_E\in\mathcal{C}(X^+,\mathbb{R}^+)\qmbox{,}K_E(\underline{x}):=
\frac{h_{\omega_0}(\underline{x})}{\omega_0(1-e^{-\omega_0})
\int\limits_{X^+}T_E h_{\omega_0}\, d\nu_{\omega_0}}$$
being defined with the help of (\ref{normal}).
\end{prop}
\pr 
$\bullet$
We first suppose that $T_E$ is integer-valued, but the image $T_E(X^+)$ is not contained 
in a proper subgroup $n\mathbb{Z}$, $n>1$ of $\mathbb{Z}$. 
Then the piecewise constant map 
$t\mapsto \kappa^{\underline{x}}_{\mathcal{I}_E}(t)$ has jumps
only at $t\in\mathbb{Z}$. We claim that even
\beq
\kappa^{\underline{x}}_{\mathcal{I}_E}(t)\sim K_E(\underline{x}) 
e^{-\omega_0 \lfloor t\rfloor}\qquad (t\rightarrow\infty),
\Leq{kappa:K}
and it is sufficient to check (\ref{kappa:K}) for $t\in\bN$.
By Lemma \ref{lem:renewal_bounds} the Fourier-Laplace transform
$$\hat{\kappa}_{\mathcal{I}_E}^{\underline{x}}(\omega)
:=\sum_{t\in\mathbb{Z}}\kappa^{\underline{x}}_{\mathcal{I}_E}(t)e^{\omega
t}\qquad (\underline{x}\in X^+)$$
of $\kappa^{\underline{x}}_{\mathcal{I}_E}$ 
converges absolutely in a strip 
${\rm Re}(\omega)\in\big(0,\omega^-(E)\big)$ of the complex plane,
with $\omega^-(E)>0$ defined in (\ref{eq:varepsilon}). 
Namely for these $\omega$ one has, with $C>0$ from 
Lemma \ref{lem:renewal_bounds},
$$\sum_{t\in\bZ\setminus \bN}\kappa^{\underline{x}}_{\mathcal{I}_E}(t)e^{\omega t}
=\kappa^{\underline{x}}_{\mathcal{I}_E}(0)\frac{1}{1-e^{-\omega}}
\qquad\mbox{and}\qquad 
\sum_{t\in\bN}\l|\kappa^{\underline{x}}_{\mathcal{I}_E}(t)e^{\omega
t}\ri|\leq \frac{C}{1-
\exp\big({\rm Re}(\omega)-\omega^-(E)\big)}.$$
The Fourier-Laplace transform of the renewal equation
(\ref{eq:renewal_equation}) is given by
$$\hat{\kappa}_{\mathcal{I}_E}^{\underline{x}}(\omega)=\frac{1}{1-e^{-\omega}}
+\sum_{\underline{y}\in\sigma^{-1}(\underline{x})}
\exp\big(\omega T_E(\underline{y})-F_E(\underline{y})\big)\ \hat{\kappa}_{\mathcal{I}_E}^{\underline{y}}(\omega)\qquad
(\underline{x}\in X^+).$$ 
In terms of the Ruelle transfer operator (\ref{eq:transfer_operator})
this leads to the formula 
\beq
\hat{\kappa}_{\mathcal{I}_E}^{\underline{x}}(\omega)
=(1-e^{-\omega})^{-1}(\idty-\mathcal{L}_{\omega})^{-1}1(\underline{x})\qquad
(\underline{x}\in X^+).
\Leq{eq:renewal_transfer_operator}
The right hand side of Eq.\ (\ref{eq:renewal_transfer_operator}) is
real-analytic in the extended strip $\mbox{Re}(\omega)\in(0,\omega_0)$, with
$\lambda_{\omega_0}=1$ from Lemma \ref{lem:unique}.

Similar as in Prop.\ 7.2 of Lalley \cite{lalley} one decomposes the 
PF-operator in the form
$$\mathcal{L}_{\omega}=\lambda_\omega\nu_\omega(\cdot)h_\omega+\mathcal{L}''_{\omega}$$ 
such that (with (\ref{normal}))
$\mathcal{L}''_{\omega}$ maps $\mathcal{C}(X^+,\mathbb{C})$
to the subspace $\{g\in\mathcal{C}(X^+,\mathbb{C}):
\nu_{k}(g)=0\}$. Thus
$$\sum_{m=0}^\infty \mathcal{L}_{\omega}^m
= \left(\sum_{m=0}^\infty \lambda_{\omega}^m\right)\nu_\omega(\cdot)h_\omega
\ +\  \sum_{m=0}^\infty (\mathcal{L}''_\omega)^m$$
and we obtain in some punctured neighbourhood of $\omega_0$
\beq\label{eq:resolvent}
(\idty-\mathcal{L}_{\omega})^{-1}=(1-\lambda_\omega)^{-1}\nu_\omega(\cdot) h_\omega
\ +\ (\idty-\mathcal{L}''_\omega)^{-1},
\eeq
the second term of the
right hand side being holomorphic. Combining Equations
(\ref{eq:renewal_transfer_operator}), (\ref{eq:resolvent}) and 
(\ref{eq:diff_lambda}) and using $\lambda_{\omega_0}=1$ we
see that the residue of 
$\omega\mapsto \hat{\kappa}_{\mathcal{I}_E}^{\underline{x}}(\omega)$ 
at $\omega=\omega_0$ equals
\beq
(1-e^{-\omega_0})^{-1}
\left(-\frac{d\lambda_{\omega}}{d \omega} (\omega_0) \right)^{-1}
 h_{\omega_0}(\underline{x})
= -(1-e^{-\omega_0})^{-1} \frac{h_{\omega_0}(\underline{x})}{\nu_{\omega_0}(T_Eh_{\omega_0})}
= -K_E(\underline{x}).
\Leq{eq:residue}
To show (\ref{kappa:K})
we introduce, similar to the proof of Thm.\ 2 of
\cite{lalley}, the function
$$F(z,\underline{x}):=\sum_{m=0}^\infty z^m 
\big(e^{m\omega_0}\kappa^{\underline{x}}_{\mathcal{I}_E}(m)-K_E(\underline{x})\big)\qquad (\underline{x}\in X^+).$$
By Lemma \ref{lem:renewal_bounds} the function $z\mapsto F(z,\underline{x})$ is
holomorphic in an open disk around $0$ of radius 
$\exp({\omega^-(E)-\omega_0})\le 1$ (see Lemma \ref{lem:unique}).
For $z$ in the open annulus $e^{-\omega_0}< |z|<e^{\omega^-(E)-\omega_0}$ we rewrite
$$F(z,\underline{x})=
\hat{\kappa}^{\underline{x}}_{\mathcal{I}_E}(\ln z+\omega_0)
\ +\ \frac{K_E(\underline{x})}{z-1}
\ -\ \sum_{\ell=1}^{\infty} z^{-\ell} e^{{-\ell}\omega_0}\kappa^{\underline{x}}_{\mathcal{I}_E}({-\ell}).$$
By Lemma \ref{lem:renewal_bounds} the last term is analytic in $z$ 
for $|z|>e^{-\omega_0}$ , whereas the  residue of the sum of the first
and second term vanishes by (\ref{eq:residue}).
Thus one can analytically extend as in \cite{lalley} the function
$z\mapsto F(z,\underline{x})$ to an open disk 
$|z|< 1+2\varepsilon(E)$ for some $\varepsilon(E)>0$.

Then Cauchy's integral formula gives
$$e^{m\omega_0}\kappa^{\underline{x}}_{\mathcal{I}_E}(m)-K_E(\underline{x})
=(2\pi i)^{-1}\int_{|z|=1+\tilde\mu(E)}F(z,\underline{x})z^{-m-1}\,dz
=\mathcal{O}\big((1+\varepsilon(E))^{-m}\big)$$
and thus $\kappa^{\underline{x}}_{\mathcal{I}_E}(m)\sim e^{-m \omega_0}K_E(\underline{x})$ 
for non-negative integer $m$.
This shows the claim for the case when $T_E$ is integer-valued and 
not contained in a proper subgroup of $\mathbb{Z}$.\\
$\bullet$
The general \emph{lattice case}, i.e.\ the image $T_E(X^+)$ generates the discrete subgroup $c\mathbb{Z}$, $c>0$ of $\mathbb{R}$
is treated by a multiplication of $T_E$ by the factor $1/c$.\\
$\bullet$
The non-lattice case follows by appropriate modification of the proof 
of Thm.\ 1 in \cite{lalley},
as the above lattice case followed by modifying the
proof of Thm.\ 2 in \cite{lalley}. \eop

The Lemmata \ref{lem:infty:empty} and \ref{lem:kappa_IE_x}, together 
with this proposition imply Part (i) of the Main Theorem:
$$\kappa_{E}^\infty(t)\asymp \kappa_{\mathcal{I}_E}(t)
\asymp\kappa_{\mathcal{I}_{E}}^{\underline{x}}(t)
\asymp\exp\big(-\omega_0(E) t\big)\qquad (E>\Eth).$$
%
\subsection{Proof of Part (ii)}
%
The escape rate $\er$ from (\ref{eq:de:escape_rate})
has been shown to equal the solution $\omega_0(E)$ of
the implicit eigenvalue formula 
$\lambda_{\rm PF}(\mathcal{L}_{\omega_0,E})=1$ 
of the transfer operator
$\mathcal{L}_{\omega,E}$ defined in (\ref{eq:transfer_operator}).

To obtain a finite-dimensional approximation of 
$\mathcal{L}_{\omega,E}$, we approximate the 
functions $F_E, T_E$,
originally defined in 
(\ref{P:T}) and (\ref{FE})
and later considered as elements of
$\mathcal{F}_{\alpha}(X^+,\mathbb{R})$,
using the homeomorphism (\ref{symb:homeo}),
by
the matrices $\widetilde{T}_E$ and $\widetilde{F}_E$ from 
(\ref{eq:approximations}), depending only on two symbols
$(k_0,k_1)$, but also considered as (locally constant) functions in $\mathcal{F}_{\alpha}(X^+,\mathbb{R})$.

It follows from (\ref{Tx}) and Lemma 10.6 of \cite{knauf} that
that there exists a $C>0$, so that with
$\delta_T(E) := C E^{-3/2}$ and $\delta_F(E) := C E^{-1}$ for
all $E>\Eth$
\begin{eqnarray}
\widetilde{T}_E^-:= \widetilde{T}_E-\delta_T(E)
\leq  & T_E  & \leq \widetilde{T}_E+\delta_T(E)=:\widetilde{T}_E^+
\qmbox { and }\label{eq:esti1_1} \\
\widetilde{F}_E^-:=\widetilde{F}_E+2\ln(1-\delta_{F}(E)) \leq &
{F}_E & \leq \widetilde{F}_E+2\ln(1+\delta_{F}(E))=:\widetilde{F}_E^+.\label{eq:esti1_2}
\end{eqnarray}

Next we define for $\vec{\delta}=(\delta_1,\delta_2)\in\bR^2$ 
and $\mathcal{M}_E$ from (\ref{M:E}) the weighted transfer matrices
$$\mathcal{M}_E(\beta,\vec{\delta}\,)\in \mathbb{R}^{\mathcal{A}\times \mathcal{A}}
\qmbox{,}
\mathcal{M}_E(\beta,\vec{\delta}\,)_{k_0,k_1}
:= \exp(\beta \delta_1+\delta_2)\ \mathcal{M}_E(\beta)_{k_0,k_1}.$$
Note that $\mathcal{M}_E^2(\beta,\vec{\delta}\,)$ has strictly positive entries, i.e.\
$\mathcal{M}_E(\beta,\vec{\delta}\,)$ is a Perron-Frobenius matrix.

With $\lambda_{\rm PF}(\mathcal{M}_E(\beta,\vec{\delta}\,))$ we denote the Perron-Frobenius eigenvalue
of the matrix $\mathcal{M}_E(\beta,\vec{\delta}\,)$.
The \emph{approximate escape rate} $\aer$ 
and its bounds $\aer^\pm$ are defined implicitly by
\beq
\lambda_{\rm PF}\Big(\mathcal{M}_E\big(\tilde{\beta}_E\big)\Big) = 1
\qmbox{ and }
\lambda_{\rm PF}
\Big(\mathcal{M}_E\big(\tilde{\beta}_E^\pm,(\pm\delta_F(E),\mp\delta_T(E))\big)\Big)=1.
\Leq{implicit}

The following lemma tells that $\aer$ and $\aer^\pm$ are well defined for large enough energies $E$:
\begin{lem}
For all $\vec{\delta}\in\bR^2$ the map  
$\beta\mapsto\lambda_{\rm PF}\big(\mathcal{M}_E(\beta,\vec{\delta}\,)\big)$ 
is continuous.  
If $\delta_2>- \inf(\widetilde{T}_E)$, it is strictly monotone increasing.
\end{lem}
\pr 
It is well known that $\lambda_{\rm PF}$ is 
related to the topological pressure by
$$\ln\l(\lambda_{\rm PF}\big(\mathcal{M}_E(\beta, \vec{\delta}\,)\big)\ri)
=P_{\rm top}\l(X,\sigma,-\widetilde{F}_E+\delta_1+\beta(\widetilde{T}_E+\delta_2)\ri).$$
Continuity of $\beta\mapsto\ln\big(\lambda_{\rm PF}(\mathcal{M}_E(\beta,\vec{\delta}\,))\big)$ follows from Theorem 9.7 of \cite{walters}.
By using the variational principle for the topological pressure and the fact that $\widetilde{T}_E+\delta_2>0$ one gets
$$\frac{\ln\big(\lambda_{\rm PF}(\mathcal{M}_E(\beta_2,\vec{\delta}\,))\big)
       -\ln\big(\lambda_{\rm PF}(\mathcal{M}_E(\beta_1,\vec{\delta}\,))\big)}
       {\beta_2-\beta_1}
\geq \inf(\widetilde{T}_E)+\delta_2 \qquad (\beta_2>\beta_1).\qquad \Box$$

\subsubsection{High Energy-Limit of the Approximate Escape Rate}
%
With $d_{\max}$ being
the maximum of the distances of the $n$ centres, we denote the
approximation to the escape rate, 
appearing in Part (ii) of the Main Theorem, by
$\beta^\infty_E:=\frac{2\sqrt{2E}\ln E}{d_{\max}}.$
\begin{lem}
$\tilde{\beta}_E$ from (\ref{implicit}) satisfies 
$\lim_{E\rightarrow\infty}\frac{\tilde{\beta}_E}{\beta_E^\infty}= 1.$
\end{lem}
\pr 
$\bullet$
First note that the estimate \
$\max_{i\in\mathcal{A}}\;\sum_{j\in\mathcal{A}}
\left(\mathcal{M}_E\big(\tilde{\beta}_E\big)\right)_{i,j}
\;\geq \;\lambda_{\rm PF}(\mathcal{M}_E\big(\tilde{\beta}_E\big))\;= \;1$\\  
implies that not all entries of the PF-matrix 
$\mathcal{M}_E\big(\tilde{\beta}_E\big)$ 
can tend to zero for $E\rightarrow\infty$.\\
This, together with the asymptotic formula
$$(\cM_E\big(\tilde{\beta}_E\big))_{k_0,k_1}\sim f^{-2}(k_0,k_1)\exp\left(-2\ln E +\tilde{\beta}_E
\Big(\frac{\overline{d}_{k_0,k_1}}{\sqrt{2E}}+\frac{Z_{k_0,k_1}\ln E}{(2E)^{3/2}}
\Big)\right)\qquad (k_0,k_1\in \mathcal{A})$$ 
for the non-zero entries of $\mathcal{M}_E\big(\tilde{\beta}_E\big)$, implies 
$$\liminf_{E\rightarrow\infty}\frac{\tilde{\beta}_E}{\beta_E^\infty}\geq 1.$$
$\bullet$
Next assume that
$$\limsup_{E\rightarrow\infty}\frac{\tilde{\beta}_E}{\beta_E^\infty}>1$$
such that at least one entry of $\mathcal{M}_E\big(\tilde{\beta}_E\big)$ is unbounded for $E\rightarrow\infty$. 
The symmetry 
$$(\mathcal{M}_E)_{(i,j),(j,k)}=(\mathcal{M}_E)_{(k,j),(j,i)}$$ 
implies that for two centres $i_0,j_0\in\{1,\ldots, n\}$ of maximal distance 
$d_{i_0,j_0}=d_{\max}$ the entry $(\mathcal{M}_E)_{k_0,k'_0}=
(\mathcal{M}_E)_{k'_0,k_0}\rightarrow\infty $ for $E\rightarrow\infty$, 
$k_0=(i_0,j_0)\in\mathcal{A}$ and $k'_0=(j_0,i_0)\in\mathcal{A}$. Then the 
$k_0$-th diagonal element of
$\l(\mathcal{M}_E\big(\tilde{\beta}_E\big)\ri)^2$ is 
unbounded as $E\to \infty$, since
$$(\mathcal{M}_E^2)_{k_0,k_0}
=\sum_{\ell\in \mathcal{A}}(\mathcal{M}_E)_{k_0, \ell} 
(\mathcal{M}_E)_{\ell,k_0}\geq (\mathcal{M}_E)_{k_0, k'_0}\cdot (\mathcal{M}_E)_{k'_0,k_0}
=(\mathcal{M}_E)_{k_0, k'_0}^2.$$
This is a contradiction to the fact that for the matrix $\mathcal{M}_E^2$ the 
diagonal elements has to be bounded above by one, as 
$\l(\mathcal{M}_E\big(\tilde{\beta}_E\big)\ri)^2$ is a 
PF-matrix whose PF-eigenvalue equals one.\\ 
Thus it follows $\lim\limits_{E\rightarrow\infty}\frac{\tilde{\beta}_E}
{\beta_E^\infty}=1$.
\hfill $\Box$
\subsubsection{Quality of the Approximation}
%
We will now estimate the quality of the approximation of $\beta_E$ by $\tilde{\beta}_E$. Recall that
the escape rate $\er$ was defined by $\lambda_{\rm PF}(\mathcal{L}_{-F_E+\beta_E T_E})=1$.
The estimates (\ref{eq:esti1_1}) and
 (\ref{eq:esti1_2}) for the functions $\widetilde{T}_E^\pm$ and $\widetilde{F}_E^\pm$ with the original
 functions $F_E$ and $T_E$ together with the monotonicity of topological pressure show that $\tilde{\beta}_E^-\leq \beta_E\leq
\tilde{\beta}_E^+$.

Together with our last lemma this completes the proof of the Main Theorem, Part (ii).
\begin{lem}
$\aer^\pm = \tilde{\beta}_E\left(1+\mathcal{O}(1/E)\right).$
\end{lem}
\pr
We are now going to express the bounds $\tilde{\beta}_E^\pm$ for the
escape rate $\beta_E$ in terms of the approximate escape rate
$\tilde{\beta}_E$. For this consider the Taylor expansion
$$\beta(\vec{\delta}\,)=\beta(0)\LA d_{\vec \delta}\beta(0),{\vec \delta}\RA+\mathcal{O}\l(\|{\vec \delta}\|^{\,2}\ri)$$
of  $\beta(\vec{\delta}\,)$,
implicitly defined by
$1=\lambda_{\rm
PF}\l(\mathcal{M}_E(\beta(\vec{\delta}\,),\vec{\delta}\,)\ri)
=:\lambda_{\rm PF}(\beta(\vec{\delta}\,),\vec{\delta}\,)$.

This implicit definition of  $\beta$ gives
$$0= d_{\vec{\delta}\,}\lambda_{\rm PF}(\beta(\vec{\delta}\,),\vec{\delta}\,)
=\l(\partial_{\beta}\lambda_{\rm PF}(\beta,\vec{\delta}\,)\ri)
d \beta(\vec{\delta}\,) 
\ +\ 
\partial_{\vec{\delta}\,}\lambda_{\rm PF}(\beta(\vec{\delta}),\vec{\delta}\,)\qquad
.$$ 
By taking the derivative of the formula
$$\lambda_{\rm PF}(\mathcal{M}_E(\beta,\vec{\delta}\,))
=\LA{\bf v}_{\rm PF}^l(\beta,{\vec \delta}\,)\,,\,\mathcal{M}_E(\beta,{\vec \delta}\,)\,{\bf v}_{\rm PF}^r(\beta,{\vec \delta}\,)\RA
\qmbox{with}
\LA{\bf v}_{\rm PF}^l(\beta,{\vec \delta}\,),{\bf v}_{\rm PF}^r(\beta,{\vec \delta}\,)\RA=1$$
it follows
$$\partial_{\delta_i}\beta(\delta_1,\delta_2)
=\frac{-\partial_{\delta_i}\lambda_{\rm PF}(\beta,{\vec \delta}\,) }
      {\partial_{\beta}\lambda_{\rm PF}(\beta,{\vec \delta}\,)}
=-\frac{\LA{\bf v}_{\rm PF}^l(\beta,{\vec \delta}\,)\,,\,\partial_{\delta_i}\mathcal{M}_E(\beta,{\vec \delta}\,)\,{\bf v}_{\rm PF}^r(\beta,{\vec \delta}\,)\RA}
       {\LA{\bf v}_{\rm PF}^l(\beta,{\vec \delta}\,)\,,\,\partial_{\beta} \mathcal{M}_E(\beta,{\vec \delta}\,)\,{\bf v}_{\rm PF}^r(\beta,{\vec \delta}\,)\RA}
       \qquad(i=1,2).$$
With the $\mathcal{A}\times
\mathcal{A}$ matrix $\widetilde{T}_E$ from (\ref{eq:approximations})
we get, $*$ denoting  the pointwise product,
\begin{eqnarray*}
\partial_{\beta} \mathcal{M}_E(\beta,\vec{\delta}\,)_{\big|(\beta({\bf 0}),{\bf 0})}
&=&\widetilde{T}_E*\mathcal{M}_E(\beta({\bf 0}))\ ,
\end{eqnarray*}
$$\partial_{\delta_1} M(\beta,{\vec\delta}\,)_{\big|(\beta({\bf 0}),{\bf 0})} 
= \mathcal{M}_E(\beta({\bf 0}))\qmbox{and}
\partial_{\delta_2} M(\beta,{\vec\delta}\,)_{\big|(\beta({\bf 0}),{\bf 0})} 
= \beta({\bf 0}) \mathcal{M}_E(\beta({\bf 0})).$$

Setting $\tau_E:={\langle {\bf v}_{\rm PF}^l(\beta({\bf 0}),{\bf 0}), 
( \widetilde{T}_E *  \mathcal{M}_E(\beta({\bf 0})){\bf
v}_{\rm PF}^r(\beta({\bf 0}),{\bf 0})\rangle}$ one gets
$$
\partial_{\delta_1}\beta(\vec{\delta}\,)_{|({\bf 0})} = -1/\tau_E\qmbox{,}
\partial_{\delta_2}\beta(\vec{\delta}\,)_{|({\bf 0})} = -\beta({\bf 0})/\tau_E
$$
and finally
$\beta(\vec{\delta}\,)=
\beta({\bf 0})-\frac{1}{\tau_E}\left(\delta_1+\beta({\bf 0})\delta_2\right)+\mathcal{O}\big(\|\vec{\delta}\|^2\big)$.
As the non-zero entries of $\widetilde{T}_E$ are bounded below by
$C/\sqrt{E}$, we get
$\tau_E^{-1}\in \mathcal{O}(E^{1/2})$.
 From  $\delta_F(E)\in \mathcal{O}(E^{-1})$ and $\delta_T(E)\in \mathcal{O}(E^{-3/2})$
(see (\ref{eq:esti1_1}) and (\ref{eq:esti1_2})) and $\beta({\bf 0})=\aer\asymp2\sqrt{2E}\ln(E)/d_{\max}$
we get the approximation
$\aer^\pm=\tilde{\beta}_E\left(1+\mathcal{O}(1/E)\right).
$\hfill$\Box$
%
\appendix
%
\section{Proof of Proposition \ref{prop:small}}
%
We use the following estimate from \cite{knauf}, Thm.\ 6.5:
If $E>\Eth$ and $\pqs\equiv x_0\in\Sigma_E$ with 
$q_0:= \|\q_0\| \geq \Rvir(E)$ and $\pm\LA\q_0,\p_0\RA \geq 0$,
then (with the symbol $\cO$ meaning existence of a bound 
for $x_\infty:=(\p_\infty,\q_\infty):=\Opm\pqs$
\beq
\p_\infty-\p_0=
\cO\l(q_0^{-1-\vep}E^{-\eh}\ri)\qmbox{,}
\q_\infty-\q_0=
\cO\l(q_0^{-\vep} E^{-1}\ri).
\Leq{DMoe:E}
From (\ref{DMoe:E}) and $\p_\infty=\cO(\sqrt{E})$ we conclude that in the case 
$\LA\q_0,\p_0\RA = 0$
\beq
\LA\q_\infty,\p_\infty\RA 
= \LA \q_\infty-\q_0 , \p_\infty \RA + \LA \q_\infty , \p_\infty-\p_0 \RA 
= \cO\l( q_0^{-\vep} E^{-\eh} \ri).
\Leq{delta:dil}
For the Kepler flow (\ref{flow:infty}), on the other hand we use the 
Lagrange-Jacobi equation, followed by an inequality valid for {\em all}
$\Zi\in\bR$:
\beq
\frac{d}{dt}\LA\q_\infty(t),\p_\infty(t)\RA =
2E+\frac{\Zi}{\|\q_\infty(t)\|}\ge E > \Eth \ge 0. 
\Leq{LJ}
The time $t_0$
needed to reach the pericentre of the Kepler flow is thus
uniquely defined by  
$\LA \q_\infty(t_0) , \p_\infty(t_0) \RA = 0$.
Together with (\ref{delta:dil}) this shows that $t_0$
is estimated by 
\beq
|t_0|\le \frac{|\LA\q_\infty,\p_\infty\RA|}{E} = 
\cO\l( q_0^{-\vep} E^{-3/2} \ri).
\Leq{est:t0}
\begin{enumerate}
\item
(\ref{est:t0}) implies Assertion 1.:\\
If the $\Phi$-scattering orbit has a
pericentre, whose distance from the origin is
larger than $\Rvir$, then by the virial inequality (\ref{VI})
that pericentre
is unique, and the total time delay is smaller than  $2|t_0|$, with
$t_0$ from (\ref{est:t0}).\\
Otherwise it enters the interaction zone at a unique time, which we
assume to equal zero w.l.o.g..
At that moment $q_0=\Rvir$ and $\LA\q_0,\p_0\RA \le 0$ and
so by (\ref{DMoe:E})
$$\q_\infty\le \Rvir+ \cO\l( \Rvir^{-\vep} E^{-1} \ri)\le 2\Rvir
$$
for $\Eth$ large enough. With (\ref{LJ}) we get that the time spent by
the Kepler orbit inside the interaction zone is smaller than
$2\Rvir/\sqrt{E}$.
\item
If $\tau_E(x)\ge t$ for $t\in(0,C_3/E^{3/2})$, then by (\ref{est:t0})
the $\Phi$-orbit through $x$ cannot have a pericentre $(\p_0,\q_0)$
with $q_0>C (t E^{3/2})^{-1/\vep}$ (here $C_3:= (C/\Rvir)^{\vep}$ is chosen so that
$q_0\ge \Rvir$). 
This implies Assertion 2, since the total symplectic volume 
of the Poincar\'e surface 
$\{(\p,\q)\in\Sigma_E \::\; \|\q\|=q_0,\LA\q,\p\,\RA<0 \}$
is of order $q_0^2E$.

\item
By our choice of $C_3$ in Assertion 2, 
$q_0 \le C \, \big(\tau_E(x) E^{3/2}\big)^{-1/\vep} \le 
C\, C_3^{-1/\vep}=\Rvir$.
\hfill $\Box$
\end{enumerate}

\end{document}